\documentclass[11pt]{article}

\usepackage[T1]{fontenc}
\usepackage[utf8]{inputenc}
\usepackage{lmodern}
\usepackage{microtype}

\usepackage{amsmath,amssymb}
\usepackage{siunitx}

\usepackage{graphicx}
\usepackage{booktabs}
\usepackage[font=footnotesize,labelfont=bf]{caption}
\usepackage{subcaption}

\usepackage[top=0.5in,bottom=0.75in,left=0.75in,right=0.75in]{geometry}
\usepackage{setspace}

\usepackage{authblk}

\usepackage[numbers,sort&compress]{natbib}
\usepackage[resetlabels]{multibib}
\newcites{S}{Supporting References}

\usepackage[colorlinks=true,linkcolor=blue,citecolor=blue,urlcolor=blue]{hyperref}

\title{\textbf{Full-Wave Harmonic Balance Framework for Dispersive Time-Varying Photonic Structures}}

\author[1]{Mohammad R. Tavakol}
\author[1,2,*]{Wenshan Cai}
\affil[1]{School of Electrical and Computer Engineering, Georgia Institute of Technology, Atlanta, Georgia 30332, USA}
\affil[2]{School of Materials Science and Engineering, Georgia Institute of Technology, Atlanta, Georgia 30332, USA}
\affil[*]{Email: wcai@gatech.edu}

\date{} 

\begin{document}

\maketitle


\begin{abstract}
\noindent
Time-varying photonic structures redistribute electromagnetic energy among Floquet harmonics, enabling frequency conversion, nonreciprocity, parametric gain, and dynamic wavefront control. Accurate modeling of realistic platforms remains challenging when temporal modulation occurs in strongly dispersive materials, because each harmonic experiences a distinct material response while remaining coupled to all others through the modulation. This work introduces a full-wave harmonic-balance framework for dispersive time-varying photonic structures. The formulation solves the steady-state Floquet response in the frequency domain by representing modulated components as induced secondary sources: volumetric polarization densities for bulk media and surface current densities for conductive sheets. Material dispersion and radiation operators are evaluated at each harmonic frequency, whereas temporal modulation enters as off-diagonal convolutional coupling in Floquet space. The framework is validated for a parametrically pumped rolled graphene cylinder, where numerical results agree with an analytical Floquet scattering solution and resolve harmonic-specific scattering, absorption, and near fields. It is further applied to an ITO-based epsilon-near-zero (ENZ) space-time metasurface, enabling shape optimization of a reflective device that converts an optical input into sidebands and redirects them into selected spatial diffraction orders. The approach establishes a general platform for modeling and designing dispersive active photonic structures with harmonic-resolved field, power, absorption, and diffraction observables.
\end{abstract}

\vspace{0.5em}
\noindent\textbf{Keywords:} Time-varying photonics, Floquet harmonics, metasurfaces, dispersion, epsilon-near-zero materials

\addtocontents{toc}{\protect\setcounter{tocdepth}{-1}}

\section{Introduction}

Time-varying photonic media extend conventional wave engineering by adding time
as an externally controlled degree of freedom. In a static structure, temporal
translation symmetry enforces conservation of optical frequency, and spatial
structuring can only redistribute the incident field among spatial modes at the
same frequency. When a material parameter is modulated in time, this symmetry is
broken: an incident wave at $\omega_0$ can exchange energy with the modulation
and generate Floquet harmonics $\omega_n=\omega_0+n\Omega$, where $\Omega$ is
the modulation frequency \cite{yin2022floquet}. If the modulation also varies in space, the generated
harmonics can simultaneously acquire transverse momentum, producing
space--time diffraction channels indexed by both temporal and spatial orders.
This additional synthetic dimension has enabled new routes to magnet-free
nonreciprocity, frequency conversion, parametric amplification, harmonic-domain
beam steering, and dynamic wavefront control
\cite{galiffi2022photonics,shaltout2015timevarying,hadad2015spacetimegradient,liu2018huygens,inampudi2019rstwa,wu2020spacetimemodulated}.

Early work on time-gradient and space--time-gradient metasurfaces showed that
properly designed temporal and transverse phase gradients can generalize Snell's
law by allowing both momentum and energy exchange with an ultrathin surface,
thereby enabling Lorentz nonreciprocity without magnetic bias
\cite{shaltout2015timevarying,hadad2015spacetimegradient}. Subsequent
metasurface platforms demonstrated that parametric sidebands are not merely
parasitic by-products of modulation, but useful output channels whose phase,
amplitude, and directionality can be engineered. Time-modulated Huygens'
metadevices used independent electric and magnetic modulation to control
sideband radiation direction and achieve efficient parametric wavefront control
\cite{liu2018huygens}, while rigorous space--time coupled-wave and free-space
$N$-path formulations provided design tools for patterned, spatially discretized,
and programmable space--time metasurfaces
\cite{inampudi2019rstwa,wu2020spacetimemodulated}. More recently, temporally
and spatiotemporally modulated metasurfaces have been used to realize optical
frequency conversion, transmissive time-modulated nonreciprocity, ultrafast
all-optical beam steering, and multifrequency beam shaping on conformal
apertures
\cite{karl2020frequencyconversion,baratisedeh2022nonreciprocity,hail2026beamsteering,pepe2026spacetimecoding}.

Beyond wavefront control, temporal modulation can change the energetic character
of a photonic system. Photonic time crystals and time-varying resonant media can
open momentum band gaps, support parametrically amplifying solutions, and convert
energy supplied by the modulation into electromagnetic radiation
\cite{ptitcyn2023floquetmie,wang2025expanding,xu2026polarization,tavakol2026nonreciprocal}.
Such effects have led to proposals and demonstrations of coherent perfect
absorption and amplification in time-varying epsilon-near-zero (ENZ) media,
all-optical polarization control through time-varying low-index films, and
nonreciprocal negative refraction enabled by photonic time crystals
\cite{galiffi2026cpa,jaffray2026polarization,tavakol2026nonreciprocal}. At the
same time, generalized space--time structures such as moving interfaces, slabs,
crystals, gradients, and wedges illustrate that even simple time-dependent
geometries can generate rich cascades of frequency-converted scattering channels
\cite{deckleger2023generalizedfdtd,bahrami2025spacetimewedges}. These advances
show that the central design variables in time-varying photonics are no longer
only geometry and static material dispersion, but also the spectrum, phase, and
spatial distribution of the modulation itself.

Accurately modeling such systems is challenging because the strongest practical
time-varying platforms are often highly dispersive. Transparent conducting
oxides, ENZ films, graphene, and resonant meta-atoms derive their large
modulation sensitivity from free-carrier or resonant material responses, whose
permittivity or conductivity changes rapidly with frequency. In these systems,
temporal modulation cannot be treated as a simple instantaneous perturbation of a
frequency-independent material parameter. Causality requires a response with
memory, and in a linear time-varying medium the polarization at time $t$ depends
on both the observation time and the earlier excitation history. This has
motivated generalized Kramers--Kronig relations, temporal-complex-polarizability
formalisms, and causality-aware descriptions of dispersive time-varying media
\cite{solis2021functional,mirmoosa2022dipole,koutserimpas2024timevarying}.
It has also exposed important numerical pitfalls: solvers that implicitly apply
time-static update rules can violate the continuity of the electric displacement
field and give incorrect results when the material parameters vary in time
\cite{mai2021timestatic}.

Several computational and semi-analytical frameworks have been developed to
address different parts of this problem. Time-domain methods, including
FDTD-based and GSTC-FDTD approaches, can represent broadband dynamics and
surface susceptibilities, but they may require long simulations to reach the
steady-state multi-harmonic response and must be implemented with care for
causal, dispersive, time-varying media
\cite{smy2020fdtd,deckleger2023generalizedfdtd}. Coupled-wave, transfer-matrix,
and modal methods provide efficient descriptions for periodic or canonical
geometries, and Floquet--Mie theory gives a rigorous solution for time-varying
dispersive spheres
\cite{inampudi2019rstwa,ptitcyn2023floquetmie}. These methods have been
instrumental in clarifying the physics of space--time diffraction, temporal
band gaps, and parametric gain. However, many practical photonic structures
combine features that are difficult to capture within a single specialized
formalism: complex two- or three-dimensional geometries, strongly dispersive
volumetric media, modulated zero-thickness sheets, multiple coupled Floquet
harmonics, and harmonic-resolved near fields, powers, and absorption.

Here, we present a full-wave harmonic-balance framework for dispersive time-varying photonic structures. The method solves directly for the steady-state Floquet response in the frequency domain, avoiding time marching while retaining the coupling among all retained harmonics. The key idea is to represent the time-varying components through induced secondary sources: volumetric polarization densities for modulated dispersive media and surface current densities for modulated material sheets. Temporal modulation then appears as a convolution in Floquet space, producing off-diagonal coupling between harmonic subproblems, whereas material dispersion, radiation operators, and source-injection operators are evaluated at the physical frequency of each harmonic. This formulation treats time-varying volumetric media and time-varying conductive sheets within the same algebraic structure, supports arbitrary temporal modulation spectra through the Fourier coefficients of the material parameters, and yields harmonic-resolved observables such as scattered power, absorption, diffraction amplitudes, and near-field distributions. This capability is particularly important for dispersive platforms, where the response at frequencies shifted from the carrier by integer multiples of the modulation frequency cannot be inferred from a single static material value at the carrier frequency. Instead, each generated harmonic must be solved with its own material response while remaining coupled to the other harmonics through the modulation-induced secondary sources. In this way, the framework combines the physical transparency of harmonic-balance and multi-frequency frequency-domain approaches \cite{Haus1991CoupledModeTheory,Shi2016MultiFrequencyFDFD} with the geometric flexibility of full-wave finite-element modeling.

We demonstrate the framework with two representative examples that establish its applicability to both surface- and volume-modulated dispersive platforms. First, we analyze a parametrically pumped rolled graphene cylinder whose Fermi level is modulated in time. This example tests the surface-current version of the formulation in a strongly dispersive THz platform and in a degenerate parametric-amplification regime where $\Omega=2\omega_0$. The harmonic-balance results are compared with an analytical Floquet scattering solution derived in the Supporting Information, showing excellent agreement for harmonic-resolved scattering and absorption. Second, we design a reflective optical space--time metasurface consisting of a binary TiO$_2$ grating on a time-modulated ITO ENZ layer backed by a gold reflector. This example tests the volumetric-polarization version of the framework in a dispersive Drude medium whose plasma frequency is modulated in time. The optimized metasurface converts the incident field into sideband harmonics and preferentially redirects the generated sidebands into the first spatial diffraction order, demonstrating frequency-converting beam deflection in a realistic optical space--time platform. By solving the coupled Floquet harmonics as a single full-wave system, the proposed harmonic-balance framework captures the essential interplay among temporal modulation, material dispersion, spatial scattering, parametric energy exchange, and harmonic-domain radiation.

\section{Harmonic-Balance Formulation}
\label{sec:HB_formulation}

\begin{figure}
\centering
\includegraphics[scale=1]{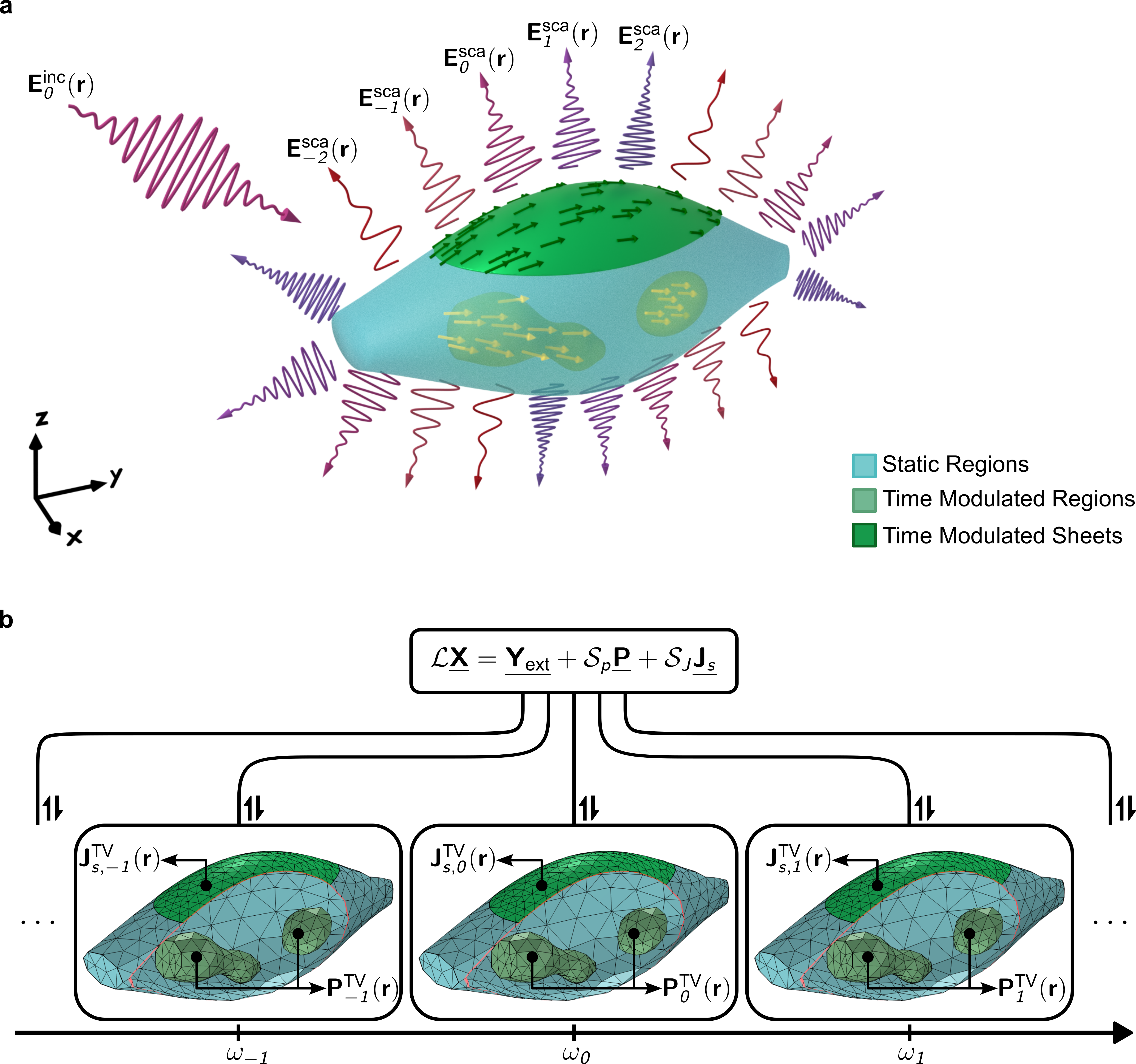}
\caption{
\textbf{Harmonic-balance solver via secondary-source representation.}
(a) Schematic of a time-varying photonic structure illuminated by an incident field $\mathbf{E}^{\rm inc}_{0}(\mathbf{r})$. Yellow vectors inside the light-green regions represent induced secondary polarization sources, while dark-green vectors on the green sheets denote induced surface-current sources. The colored outgoing waves indicate multi-harmonic scattering generated by these secondary sources. (b) Harmonic-balance representation of the same system as a set of coupled frequency-domain FEM subproblems. Each mesh view corresponds to one Floquet harmonic $\omega_n$, with the subproblems coupled through the secondary sources governed by the global multi-harmonic system.
}
\label{fig:Fig1}
\end{figure}

The proposed full-wave harmonic-balance framework is schematically summarized in
Fig.~\ref{fig:Fig1}. We consider a general dispersive photonic structure containing
static regions, time-modulated volumetric media, and time-modulated material sheets
or interfaces. As illustrated in Fig.~\hyperref[fig:Fig1]{\ref*{fig:Fig1}(a)}, the structure is illuminated
by an incident field at angular frequency $\omega_0$, while the time-varying
components generate frequency-converted scattered fields at multiple Floquet
harmonics. In this representation, the action of the time-varying elements is
described through induced secondary sources: polarization densities in modulated
volumetric regions and surface current densities on modulated sheets. These
secondary sources provide the physical mechanism by which energy is transferred
among harmonic channels.

For a modulation frequency $\Omega$, the generated harmonic frequencies are
\begin{equation}
    \omega_n = \omega_0 + n\Omega ,
    \label{eq:HB_frequencies}
\end{equation}
where $n$ is the Floquet-harmonic index. The central objective of the
harmonic-balance formulation is to solve for the steady-state electromagnetic
response at these coupled harmonic frequencies directly, without time marching.
For a time-periodic system, any field or induced source quantity can be expanded
in a truncated Floquet series as
\begin{equation}
    \mathbf{F}(\mathbf{r},t)
    =
    \sum_{n=-N}^{N}
    \mathbf{F}_{n}(\mathbf{r}) e^{j\omega_n t},
    \label{eq:HB_ansatz}
\end{equation}
where $2N+1$ harmonics are retained. The unknown electromagnetic fields are
collected into a multi-harmonic vector
\begin{equation}
    \underline{\mathbf{X}}
    =
    \begin{bmatrix}
        \mathbf{X}_{-N} &
        \cdots &
        \mathbf{X}_{0} &
        \cdots &
        \mathbf{X}_{N}
    \end{bmatrix}^{T},
    \label{eq:HB_unknown_vector}
\end{equation}
where the underline denotes the vector space of retained Floquet harmonics. Each
component $\mathbf{X}_n$ represents the field unknowns at frequency $\omega_n$.
Thus, the original time-varying problem is recast as a coupled set of
frequency-domain electromagnetic problems.

The induced polarization and surface-current sources are expanded in the same
harmonic basis,
\begin{equation}
    \underline{\mathbf{P}}
    =
    \begin{bmatrix}
        \mathbf{P}_{-N} &
        \cdots &
        \mathbf{P}_{0} &
        \cdots &
        \mathbf{P}_{N}
    \end{bmatrix}^{T},
    \qquad
    \underline{\mathbf{J}_s}
    =
    \begin{bmatrix}
        \mathbf{J}_{s,-N} &
        \cdots &
        \mathbf{J}_{s,0} &
        \cdots &
        \mathbf{J}_{s,N}
    \end{bmatrix}^{T}.
    \label{eq:HB_source_vectors}
\end{equation}
In a static linear system, the response at each frequency is independent. In
contrast, a time-periodic material coefficient mixes spectral components through
convolution in harmonic space. As a result, the field at one harmonic can induce
polarization or surface current at other harmonics, which then radiate or scatter
back into the electromagnetic system. This is the physical origin of the
multi-harmonic scattering channels shown in Fig.~\hyperref[fig:Fig1]{\ref*{fig:Fig1}(a)}.

Using this secondary-source representation, the multi-harmonic electromagnetic
problem can be written compactly as
\begin{equation}
    \mathcal{L}\,\underline{\mathbf{X}}
    =
    \underline{\mathbf{Y}_{\rm ext}}
    +
    \mathcal{S}_{P}\underline{\mathbf{P}}
    +
    \mathcal{S}_{J}\underline{\mathbf{J}_s}.
    \label{eq:HB_source_form}
\end{equation}
Here, $\mathcal{L}$ is the block frequency-domain Maxwell operator for the static
background structure, $\underline{\mathbf{Y}_{\rm ext}}$ is the external excitation
projected into the harmonic space, and $\mathcal{S}_{P}$ and $\mathcal{S}_{J}$
are source-injection operators for volumetric polarization and sheet-current
sources, respectively. Equation~\eqref{eq:HB_source_form} should not be
interpreted as an uncoupled driven problem, because the secondary sources
$\underline{\mathbf{P}}$ and $\underline{\mathbf{J}_s}$ are not prescribed
external sources. They are induced by the unknown multi-harmonic fields through
the time-varying material response. The source-injection blocks may depend on the
harmonic frequency, for example through $\mathcal{S}_{P,n}(\omega_n)$ and
$\mathcal{S}_{J,n}(\omega_n)$, so that dispersive source-to-field mappings,
radiation conditions, and sheet/source normalization factors are evaluated at the
physical frequency of each harmonic. The inter-harmonic coupling itself,
however, is contained in the source-generation relations described below.

To make the coupling explicit, consider a material or sheet parameter with a
temporal Fourier expansion
\begin{equation}
    m(\mathbf{r},t)
    =
    \sum_{\ell=-L}^{L}
    m_{\ell}(\mathbf{r}) e^{j\ell\Omega t}.
    \label{eq:HB_modulation_spectrum}
\end{equation}
Multiplication of this modulation waveform by the multi-harmonic field produces
a convolution in Floquet space. Therefore, the induced secondary sources at
$\omega_n$ generally depend on field components at neighboring or more distant
harmonics,
\begin{equation}
    \mathbf{P}_{n}
    =
    \sum_{\ell=-L}^{L}
    \mathcal{C}^{P}_{n,n-\ell}\mathbf{X}_{n-\ell},
    \qquad
    \mathbf{J}_{s,n}
    =
    \sum_{\ell=-L}^{L}
    \mathcal{C}^{J}_{n,n-\ell}\mathbf{X}_{n-\ell}.
    \label{eq:HB_explicit_coupling}
\end{equation}
Here, $\mathcal{C}^{P}_{n,n-\ell}$ and $\mathcal{C}^{J}_{n,n-\ell}$ denote the
volumetric and sheet source-generation blocks associated with the $\ell$-th
modulation order; terms outside the retained harmonic window are omitted by the
truncation. A single-tone modulation produces nearest-neighbor harmonic coupling,
whereas a multi-tone or non-sinusoidal modulation produces a block-banded
coupling matrix whose harmonic bandwidth is determined by $L$. This
representation naturally supports arbitrary modulation spectra, including
traveling-wave, multi-tone, and waveform-engineered temporal modulations.

In global form, the induced-source relation can be written as
\begin{equation}
    \underline{\mathbf{P}}
    =
    \mathcal{C}_{P}\underline{\mathbf{X}},
    \qquad
    \underline{\mathbf{J}_s}
    =
    \mathcal{C}_{J}\underline{\mathbf{X}},
    \label{eq:HB_CP_CJ_relation}
\end{equation}
or, equivalently,
\begin{equation}
    \underline{\mathbf{Q}}
    =
    \mathcal{C}\,\underline{\mathbf{X}},
    \qquad
    \underline{\mathbf{Q}}
    =
    \begin{bmatrix}
        \underline{\mathbf{P}} \\
        \underline{\mathbf{J}_s}
    \end{bmatrix},
    \qquad
    \mathcal{C}
    =
    \begin{bmatrix}
        \mathcal{C}_{P} \\
        \mathcal{C}_{J}
    \end{bmatrix}.
    \label{eq:HB_coupling_relation}
\end{equation}
The operators $\mathcal{C}_{P}$ and $\mathcal{C}_{J}$ are generally non-diagonal
in harmonic space. Their diagonal blocks describe the same-frequency material
response, while their off-diagonal blocks describe frequency conversion between
harmonics separated by integer multiples of $\Omega$. The number and strength of
these off-diagonal couplings are determined by the Fourier spectrum of the
temporal modulation.

Defining
\begin{equation}
    \mathcal{S}
    =
    \begin{bmatrix}
        \mathcal{S}_{P} & \mathcal{S}_{J}
    \end{bmatrix},
    \label{eq:HB_combined_source_operator}
\end{equation}
the secondary-source formulation becomes
\begin{equation}
    \mathcal{L}\,\underline{\mathbf{X}}
    =
    \underline{\mathbf{Y}_{\rm ext}}
    +
    \mathcal{S}\underline{\mathbf{Q}}.
    \label{eq:HB_compact_source_form}
\end{equation}
Substitution of the induced-source relations gives the closed global
harmonic-balance system
\begin{equation}
    \left(
    \mathcal{L}
    -
    \mathcal{S}_{P}\mathcal{C}_{P}
    -
    \mathcal{S}_{J}\mathcal{C}_{J}
    \right)
    \underline{\mathbf{X}}
    =
    \underline{\mathbf{Y}_{\rm ext}},
    \label{eq:HB_system_expanded}
\end{equation}
or, in compact form,
\begin{equation}
    \mathcal{A}_{\rm HB}\underline{\mathbf{X}}
    =
    \underline{\mathbf{Y}_{\rm ext}},
    \qquad
    \mathcal{A}_{\rm HB}
    =
    \mathcal{L}
    -
    \mathcal{S}\mathcal{C}.
    \label{eq:HB_system}
\end{equation}
Equation~\eqref{eq:HB_system} is the central algebraic form of the proposed
framework and corresponds to the global multi-harmonic system shown in
Fig.~\hyperref[fig:Fig1]{\ref*{fig:Fig1}(b)}. The diagonal blocks represent
independent full-wave problems at the harmonic frequencies $\omega_n$, whereas
the off-diagonal blocks originate from the secondary sources induced by the
time-varying components. Therefore, the full time-varying photonic structure is
solved as a single coupled system in the extended Floquet-frequency space.

In a finite-element implementation, each harmonic component is solved using a
frequency-domain weak form. The conventional FEM unknown vector is lifted into a
harmonic-domain vector, so that each spatial degree of freedom carries multiple
harmonic coefficients. As depicted by the mesh views in Fig.~\hyperref[fig:Fig1]{\ref*{fig:Fig1}(b)},
each harmonic subproblem retains the full spatial complexity of the electromagnetic
geometry, while remaining coupled to the other harmonics through the
secondary-source terms. For $2N+1$ retained harmonics, the total number of
unknowns scales approximately by the same factor relative to a single-frequency
FEM problem. However, the resulting matrix has a structured block form: the
diagonal blocks contain the standard full-wave FEM operators at individual
harmonic frequencies, while the off-diagonal blocks encode coupling through the
induced polarization and surface-current sources.

The present framework is closely related in spirit to multi-frequency
frequency-domain solvers developed for active nanophotonic devices. In particular,
the MF-FDFD algorithm of Shi \textit{et al.} couples multiple frequency-domain
Maxwell systems through modulation-induced polarization terms~\cite{Shi2016MultiFrequencyFDFD}.
The present formulation generalizes this viewpoint in several directions. First,
the modulation spectrum is not restricted to a single sinusoidal tone: an
arbitrary number of temporal Fourier components can be included through the
coupling operators $\mathcal{C}_{P}$ and $\mathcal{C}_{J}$. Second, by casting
the problem in a full-wave FEM setting, complex three-dimensional geometries,
curved interfaces, inhomogeneous materials, anisotropic media, and sheet-based
material models can be treated without reducing the device to a coupled-mode or
equivalent-circuit description. Third, frequency dispersion is incorporated
directly at every harmonic through $\mathcal{L}(\omega_n)$ and the
frequency-dependent source-injection/source-generation blocks, avoiding slowly
varying envelope or weak-dispersion assumptions commonly invoked in coupled-mode
treatments~\cite{Haus1991CoupledModeTheory}.

This full-wave harmonic-balance formulation is particularly suitable for
dispersive time-varying photonic structures. Since each harmonic is treated
directly in the frequency domain, material dispersion can be evaluated at the
actual generated frequency $\omega_n$. At the same time, temporal modulation is
retained through the coupling operators $\mathcal{C}_{P}$ and $\mathcal{C}_{J}$,
allowing dispersion and fast time modulation to be incorporated simultaneously.
The formulation therefore avoids the need for long transient simulations and
directly provides the steady-state amplitudes, phases, and field profiles of all
retained Floquet harmonics.

Once the global system is solved, the harmonic-domain solution can be used to
reconstruct the steady-state time-domain field,
$\mathbf{X}(\mathbf{r},t)=\sum_{n=-N}^{N}\mathbf{X}_n(\mathbf{r})
e^{j\omega_n t}$. Harmonic-resolved observables are then extracted from the
corresponding frequency-domain fields. The power carried by the $n$-th harmonic
through a surface $S$ is computed from the time-averaged Poynting flux,
\begin{equation}
    \mathcal{P}_n
    =
    \frac{1}{2}
    {\rm Re}
    \int_S
    \mathbf{E}_n(\mathbf{r})
    \times
    \mathbf{H}_n^{*}(\mathbf{r})
    \cdot
    \hat{\mathbf{n}}\,dA ,
    \label{eq:HB_poynting}
\end{equation}
where $\mathbf{H}_n$ is obtained from Faraday's law as
$\mathbf{H}_n=-[j\omega_n\mu(\omega_n)]^{-1}\nabla\times\mathbf{E}_n$.
For port-based structures, the transmission and reflection coefficients at each
harmonic are obtained by projecting the total field onto the corresponding guided
modes of the input and output port cross-sections. For scattering configurations,
harmonic-resolved scattered power, absorption, and cross sections are computed
directly from the near-field or far-field quantities at each $\omega_n$.

The accuracy of the method is controlled primarily by the harmonic truncation
order. Harmonics outside the retained set are neglected, which can introduce
truncation error if significant frequency-converted power is generated beyond the
selected range. In practice, convergence is assessed by increasing $N$ and
monitoring harmonic-resolved observables such as $\mathcal{P}_n$, scattering
coefficients, absorption, near-field distributions, and induced source amplitudes.
A typical convergence criterion is
\begin{equation}
    \frac{
    \left|
    \mathcal{P}_{n}^{(N+1)}
    -
    \mathcal{P}_{n}^{(N)}
    \right|
    }{
    \left|
    \mathcal{P}_{n}^{(N)}
    \right|
    }
    < \eta ,
    \label{eq:HB_convergence}
\end{equation}
where $\eta$ is a prescribed tolerance. Once these quantities become insensitive
to further expansion of the harmonic basis, the retained Floquet space provides
an accurate representation of the steady-state response.

Overall, the proposed harmonic-balance framework transforms a dispersive
time-varying electromagnetic problem into a coupled set of full-wave
frequency-domain subproblems. By representing the time-varying components through
induced secondary polarization and surface-current sources, the method provides a
physically transparent and computationally efficient route for analyzing
multi-harmonic scattering, absorption, and field conversion in time-modulated
photonic structures.

\section{Results and Discussion}

\subsection{Harmonic-Resolved Scattering from a Parametrically Pumped Graphene Cylinder}

\begin{figure}
\centering
\includegraphics[scale=1]{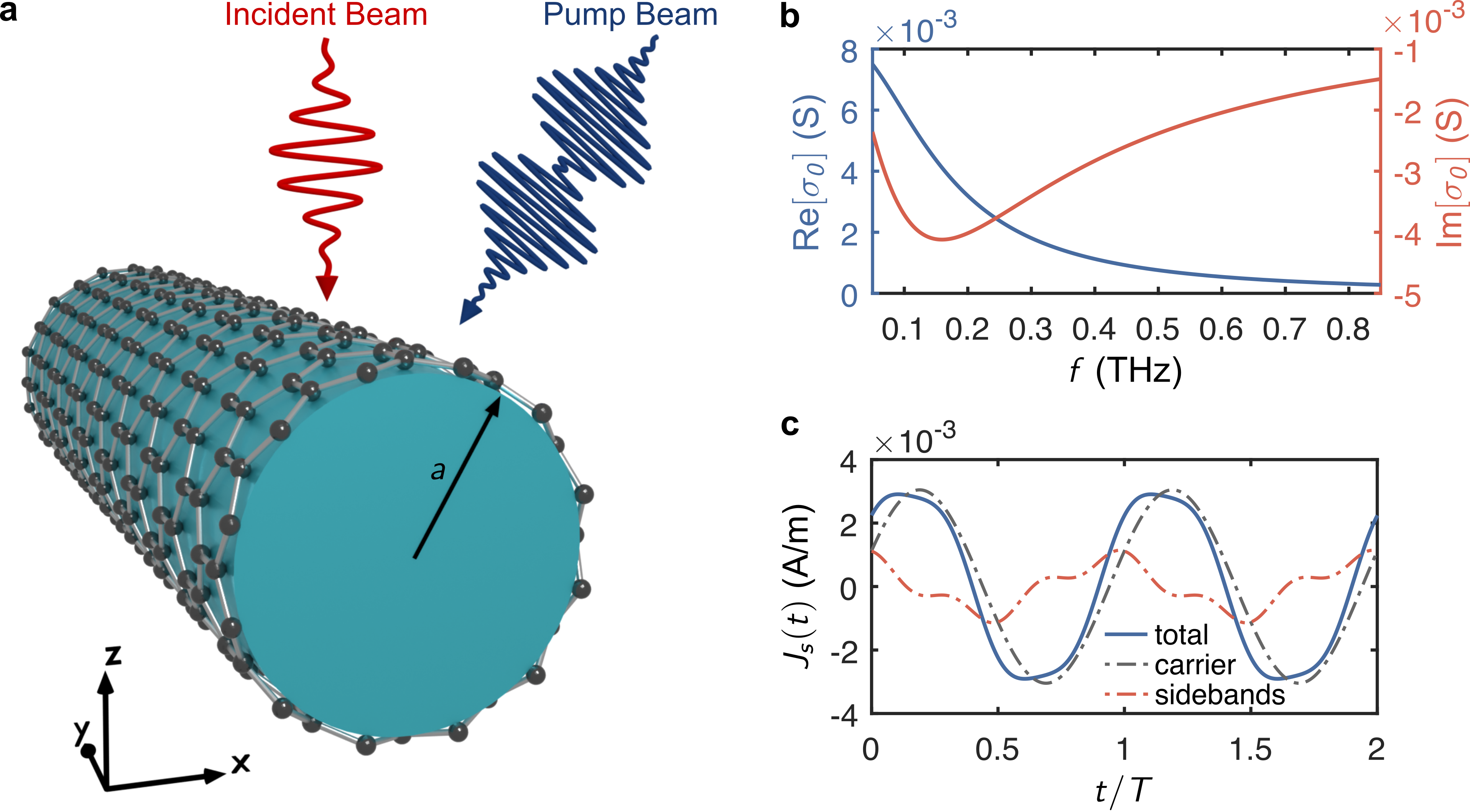}
\caption{
\textbf{Parametrically pumped rolled graphene cylinder configuration.}
(a) Schematic of a graphene-wrapped dielectric cylinder illuminated by an incident beam and driven by an external pump that modulates the graphene Fermi level. The dielectric core has $\epsilon_r=2.25$ and radius $a=40~\mu\mathrm{m}$. (b) Real and imaginary parts of the unmodulated graphene surface conductivity $\sigma_0$ for $E_{F0}=70~\mathrm{meV}$, showing the dispersive response over the THz band. (c) Steady-state surface current density $J_s(t)$ induced by a normally incident electric field with $E_0=1~\mathrm{V/m}$ and $f_0=0.4~\mathrm{THz}$. The total current is decomposed into the carrier contribution and modulation-induced sidebands, with time normalized to the incident-field period $T$.
}
\label{fig:Fig2}
\end{figure}

To demonstrate the proposed harmonic-balance framework in a strongly dispersive
and resonant time-varying surface platform, we consider the scattering response
of a dielectric cylinder wrapped by a graphene sheet with a temporally modulated
Fermi level. The structure is shown schematically in
Fig.~\hyperref[fig:Fig2]{\ref*{fig:Fig2}(a)}. The dielectric core has radius
$a=40~\mu\mathrm{m}$ and relative permittivity $\epsilon_d=2.25$, and it is coated
by a rolled graphene layer. Such a geometry is not merely a conceptual model:
recent experimental work has demonstrated graphene rolls with controllable
geometry and chirality, indicating that rolled graphene architectures are a
feasible route for realizing cylindrical graphene-based photonic structures
\cite{Zhang2025GrapheneRolls}. In addition, optical pump--THz probe experiments
have shown that the THz conductivity of graphene can be modified on picosecond
and subpicosecond time scales through photoexcitation, providing a physical basis
for optically driven temporal modulation of graphene in the THz regime
\cite{Tasolamprou2019UltrafastTHz,Tomadin2018UltrafastGraphene}.

The design starts from the corresponding static, unmodulated graphene cylinder.
Under TM illumination, the graphene coating supports localized surface-plasmon
resonances, whose spectral position can be controlled through the cylinder
radius, the dielectric permittivity of the core, and the graphene Fermi level.
The geometric and material parameters are selected such that, for an unmodulated
graphene Fermi level of $E_{F0}=70~\mathrm{meV}$, the static structure exhibits a
pronounced localized plasmonic resonance near the operating frequency
$f_0=0.4~\mathrm{THz}$. The corresponding unmodulated graphene conductivity is
shown in Fig.~\hyperref[fig:Fig2]{\ref*{fig:Fig2}(b)}, where both the real and
imaginary parts display the expected dispersive Drude response in the low-THz
range for a graphene relaxation time of $\tau=1~\mathrm{ps}$. The cylinder is
excited by a normally incident TM-polarized wave with magnetic-field amplitude
$H_y=1~\mathrm{A/m}$.
After establishing the resonant static design, we introduce temporal modulation
by photoexciting the graphene roll. The Fermi level is prescribed as
\begin{equation}
    E_F(t)
    =
    E_{F0}
    +
    \Delta E_F \cos(\Omega t+\alpha),
\end{equation}
where $\Delta E_F=40~\mathrm{meV}$. We intentionally select a fast modulation
frequency satisfying $\Omega=2\omega_0$, which places the cylinder in the
degenerate parametric-amplification regime. Under this condition, the incident
harmonic at $\omega_0$ is directly coupled to the down-converted Floquet
harmonic $\omega_{-1}=\omega_0-\Omega=-\omega_0$. This is the
temporal-scattering analogue of modulating a resonant oscillator at twice its
natural frequency, where the pump supplies energy to the signal and idler
channels. In time-modulated resonant scatterers, this condition is also
associated with the opening of a momentum band gap in the equivalent photonic
time-crystal picture, and the amplifying branch inside this gap underlies the
parametric enhancement of the scattered fields
\cite{Asadchy2022ParametricMie}. Thus, this example is deliberately
chosen as a stringent test case: it combines a resonant graphene plasmonic
response, strong material dispersion, and rapid temporal modulation at twice the
incident frequency.

The time-domain surface-current waveform of the modulated graphene sheet is
shown in Fig.~\hyperref[fig:Fig2]{\ref*{fig:Fig2}(c)}, where the graphene is
driven by a monochromatic tangential electric field with amplitude
$1~\mathrm{V/m}$. The analytical steady-state form of this time-domain graphene
surface current is derived in the Supporting Information. Even though the
excitation is monochromatic, the temporal modulation of the Fermi level generates
a multi-harmonic surface current. The total current can therefore be interpreted
as the superposition of the carrier response and modulation-induced sideband
contributions. This behavior is precisely the type of response that the proposed
harmonic-balance formulation is designed to capture: the dispersive graphene
surface current is solved simultaneously across the coupled Floquet harmonics,
rather than by treating the harmonics as independent frequency-domain problems.

\begin{figure}
\centering
\includegraphics[scale=1]{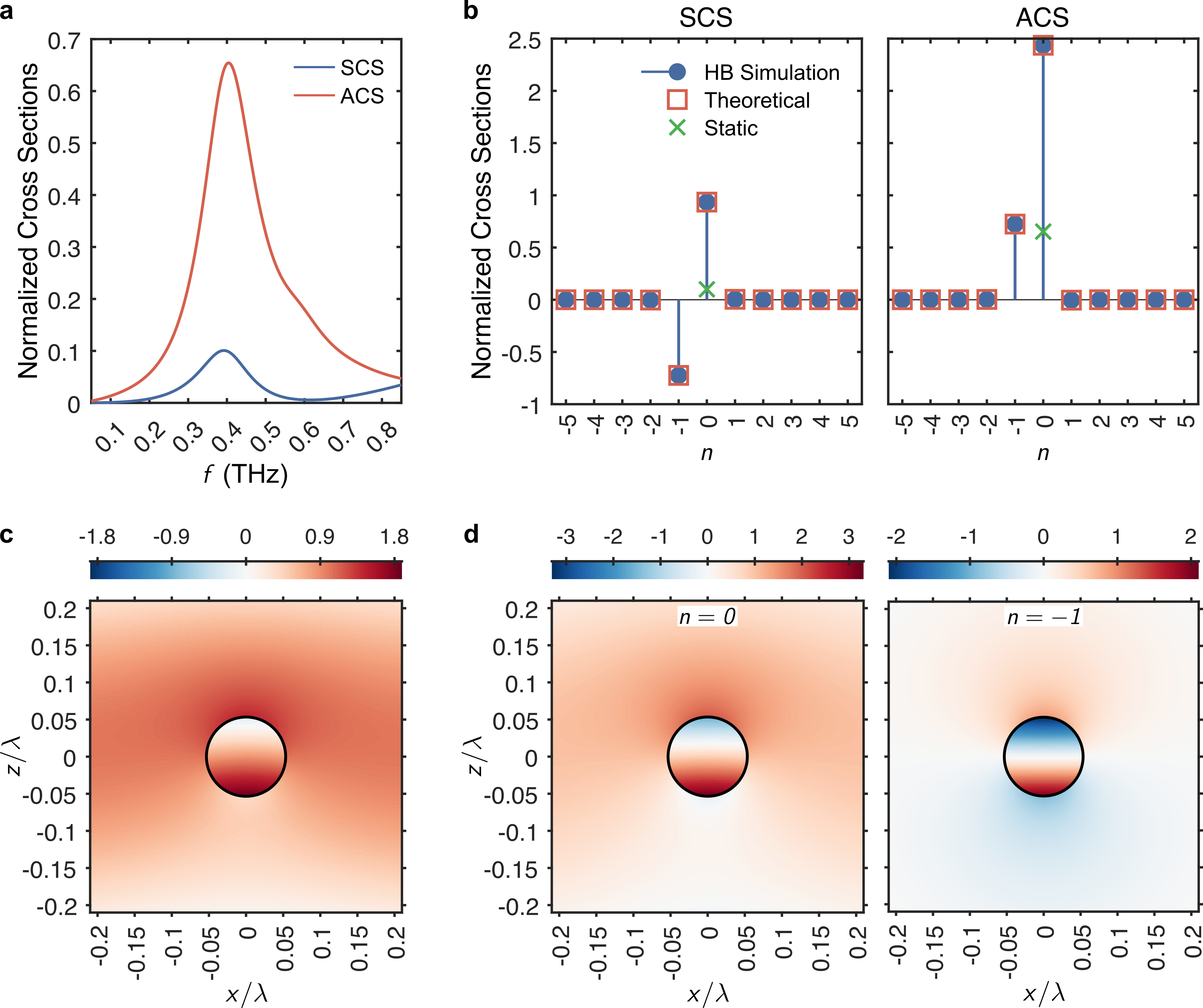}
\caption{
\textbf{Harmonic-resolved scattering from a parametrically pumped rolled graphene cylinder.}
(a) Normalized scattering and absorption cross sections of the static graphene cylinder, obtained in the absence of Fermi-level modulation, $\Delta E_F=0$. 
(b) Harmonic-resolved normalized scattering and absorption cross sections of the modulated structure under parametric pumping. The harmonic-balance FEM results are compared with the analytical solution, while the green crosses show the corresponding static response to highlight the redistribution of scattering and absorption among Floquet harmonics. 
(c) Magnetic-field distribution, $H_z$, at $f_0=0.4~\mathrm{THz}$ for the static (unmodulated) graphene cylinder. 
(d) Magnetic-field distributions of the zeroth-order and down-converted Floquet harmonics, $n=0$ and $n=-1$, for the modulated cylinder. Color bars in panels (c,d) are in units of A/m.
}
\label{fig:Fig3}
\end{figure}

Figure~\ref{fig:Fig3} presents the harmonic-resolved scattering and absorption
results. In the static case, shown in
Fig.~\hyperref[fig:Fig3]{\ref*{fig:Fig3}(a)}, the normalized scattering and
absorption cross sections, $\sigma_{\rm sca}/(2a)$ and
$\sigma_{\rm abs}/(2a)$, are confined to the incident harmonic because
$\Delta E_F=0$. The peak response at $f_0=0.4~\mathrm{THz}$ confirms that the
selected static parameters place the rolled graphene cylinder near its localized
plasmonic resonance. This static resonance provides the electromagnetic
enhancement that is subsequently leveraged by the temporal modulation.

When the Fermi level is modulated at $\Omega=2\omega_0$, the response changes
qualitatively. As shown in Fig.~\hyperref[fig:Fig3]{\ref*{fig:Fig3}(b)},
scattering and absorption are redistributed among the Floquet harmonics, with
the dominant response appearing at the carrier harmonic $n=0$ and the
down-converted harmonic $n=-1$. The normalized cross sections increase
substantially compared with the unmodulated case, indicating that energy
supplied by the pump is converted into radiating and absorbed Floquet channels.
The harmonic-balance FEM results agree closely with the analytical Floquet
scattering solution derived in the Supporting Information. This agreement is
important because the analytical solution is obtained from the direct cylindrical
Floquet formulation, whereas the numerical solution follows from the proposed
finite-element harmonic-balance framework with time-varying dispersive surface
currents. Their consistency verifies that the framework accurately captures both
the graphene dispersion and the modulation-induced coupling between harmonics.

The field maps in Fig.~\hyperref[fig:Fig3]{\ref*{fig:Fig3}(c)} and
Fig.~\hyperref[fig:Fig3]{\ref*{fig:Fig3}(d)} provide a complementary view of
this process. In the static case, the magnetic-field distribution $H_y$ at
$f_0=0.4~\mathrm{THz}$ exhibits the localized resonant field pattern associated
with the graphene-coated cylinder. Under parametric pumping, strong field
distributions are observed not only at the carrier harmonic $n=0$, but also at
the down-converted harmonic $n=-1$. The appearance of a pronounced $n=-1$ field
confirms that the time-modulated graphene sheet acts as an active
frequency-converting boundary, transferring pump energy into the Floquet
sideband. The enhanced near fields and cross sections therefore result from the
combined action of the localized graphene plasmon resonance and the parametric
coupling induced by the time-varying Fermi level.

This example highlights the central capability of the harmonic-balance
formulation developed in this work. The method resolves a realistic dispersive
time-varying surface whose modulation frequency is comparable to the signal
frequency, whose material response is strongly frequency dependent, and whose
scattering behavior is governed by coupled Floquet harmonics. The close
agreement with the analytical solution confirms the accuracy of the proposed
framework, while the parametrically pumped rolled graphene cylinder demonstrates
its applicability to resonant time-varying photonic structures exhibiting
harmonic conversion, parametric enhancement, and harmonic-resolved scattering
and absorption.

\subsection{Optical Space--Time Metasurface Based on Modulated ITO}
\label{subsec:ITO_ST_metasurface}

\begin{figure}
\centering
\includegraphics[scale=1]{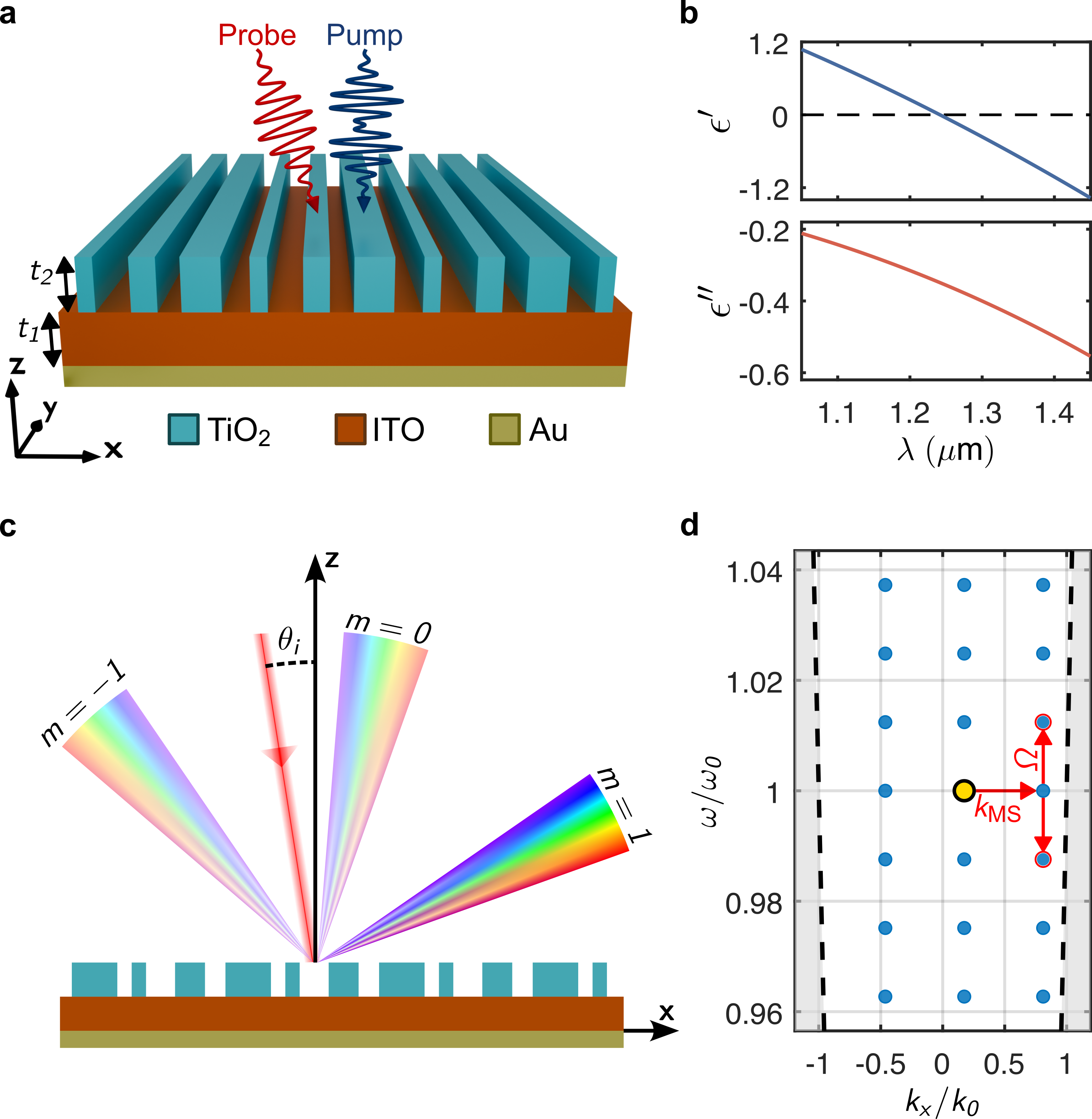}
\caption{
\textbf{Configuration of an optical space--time metasurface based on modulated ITO.}
(a) Three-dimensional schematic of the reflective space--time metasurface composed of a binary corrugated TiO$_2$ grating on a time-varying ITO layer backed by a gold reflector. The probe beam impinges on the metasurface while an external pump modulates the ITO response in time. The TiO$_2$ and ITO layer thicknesses are $t_1=300~\mathrm{nm}$ and $t_2=300~\mathrm{nm}$, respectively.
(b) Real and imaginary parts of the ITO permittivity, $\epsilon=\epsilon'-j\epsilon''$, in the epsilon-near-zero spectral region.
(c) Two-dimensional $x$--$z$ cross section illustrating space--time diffraction from the reflective metasurface, which is uniform along the $y$ direction. The incident beam arrives at $\theta^{\rm inc}=10^\circ$, and the diffracted beams correspond to different spatial orders. The bright rainbow beam denotes the targeted space--time diffraction channel, whereas the weaker beams indicate other propagating orders.
(d) Dispersion diagram of the space--time diffraction channels. The yellow marker denotes the incident wave at $\omega_0$, and the blue markers indicate propagating diffraction orders allowed by the light-cone constraint. Adjacent orders are separated by $k_{\rm MS}=2\pi/\Lambda$ along the transverse-momentum axis and by the modulation frequency $\Omega$ along the frequency axis. The red-circled channels indicate the target orders with spatial index $m=+1$, for which the metasurface is designed to enhance the diffracted amplitude.
}
\label{fig:Fig4}
\end{figure}

As a second demonstration, we apply the proposed harmonic-balance framework to a reflective optical space--time metasurface based on a dispersive time-modulated epsilon-near-zero material. The structure, shown in Fig.~\ref{fig:Fig4}, consists of a binary TiO$_2$ grating patterned on top of an ITO layer and backed by a gold reflector. In contrast to the graphene-cylinder example, where the time-varying response is introduced through a modulated surface conductivity, here the time-varying material is volumetric and dispersive. The ITO layer is modeled by a Drude response whose plasma frequency is modulated in time, representing the ultrafast pump-induced change of the free-carrier response. This configuration therefore provides a stringent test of the proposed framework because material dispersion and temporal modulation must be treated simultaneously across multiple generated harmonics.

ITO is chosen because transparent conducting oxides exhibit strong optical nonlinearities near their epsilon-near-zero wavelength, where a small change in carrier response can produce a large change in refractive index and permittivity~\cite{alam2016large}. In the present design, the unmodulated ITO parameters are selected such that the real part of the permittivity crosses zero near the operating wavelength, $\lambda_0=1240~\mathrm{nm}$, as shown in Fig.~\hyperref[fig:Fig4]{\ref*{fig:Fig4}(b)}. Operating in this ENZ regime maximizes the effect of plasma-frequency modulation on the electromagnetic response. This choice is consistent with recent experimental demonstrations of ultrafast wavefront shaping, time-varying gradient metasurfaces, and space--time optical diffraction based on pumped ITO and related transparent conducting oxides~\cite{fan2023ultrafast,karimi2023timevarying,harwood2025spacetime}.

The dispersive time-varying ITO response is incorporated through the Drude polarization relation. Denoting $A(t)=\omega_p^2(t)$, the free-carrier part of the polarization is governed in the frequency domain by a convolution between the plasma-frequency modulation and the electric field. In compact form, the total polarization is written as
\begin{equation}
    P(\omega)
    =
    \epsilon_0(\epsilon_\infty-1)E(\omega)
    -
    \frac{\epsilon_0}{\omega(\omega-j\gamma)}
    \frac{1}{2\pi}
    \left[A*E\right](\omega),
    \label{eq:ITO_Drude_convolution}
\end{equation}
where $\epsilon_\infty$ is the background permittivity, $\gamma$ is the damping rate, and $*$ denotes convolution in frequency. For a periodic modulation, this relation reduces to the discrete harmonic coupling equation
\begin{equation}
    P_n
    =
    \epsilon_0(\epsilon_\infty-1)E_n
    -
    \frac{\epsilon_0}{\omega_n(\omega_n-j\gamma)}
    \sum_k A_{n-k}E_k,
    \qquad
    \omega_n=\omega_0+n\Omega .
    \label{eq:ITO_Drude_harmonic_coupling}
\end{equation}
Thus, the Drude denominator is evaluated at each generated harmonic frequency, while the Fourier coefficients of $A(t)$ determine the coupling strength between harmonics. This is precisely the situation for which the harmonic-balance framework is designed: the dispersive material response is not approximated by a static permittivity, and the temporal modulation enters as off-diagonal coupling in Floquet space.

We assume a single-tone modulation of the ITO plasma frequency with modulation frequency $\Omega/2\pi=3~\mathrm{THz}$. Equivalently, the squared plasma frequency can be written as $A(t)=A_0[1+M_A\cos(\Omega t+\alpha)]$, where $M_A=(\omega_{p,\max}^2-\omega_{p,\min}^2)/(\omega_{p,\max}^2+\omega_{p,\min}^2)$. In the simulations, the pump-induced swing is chosen such that the pumped and unpumped plasma frequencies differ by $10\%$, with $\omega_{p,\min}=0.9\,\omega_{p,\max}$. This corresponds to a modulation index of approximately $M_A\simeq0.105$ for the squared plasma frequency. The modulation frequency is chosen close to experimentally accessible multi-terahertz beat frequencies used in optical space--time modulation; for example, Guo \textit{et al.}\ demonstrated a heterodyne-pump-induced temporal modulation near $2.8~\mathrm{THz}$ in a nonlinear optical metasurface~\cite{guo2019nonreciprocal}.

The purpose of the metasurface is to realize frequency-converting beam deflection in reflection. Therefore, this device should not be interpreted as a conventional static phase-gradient metasurface. In a static gradient metasurface, the design target is usually the phase of the reflected field at the incident frequency. Here, instead, the target channel is a space--time diffraction order: the incident wave at $\omega_0$ is converted to sideband frequencies $\omega_n=\omega_0+n\Omega$ and redirected into selected spatial diffraction orders. The diffraction angles are determined by conservation of transverse momentum and energy. For reflected space--time orders, the allowed angles satisfy $\sin(\theta_{mn})=[\sin(\theta_i)+mK/k_0]/[1+n\Omega/\omega_0]$, where $K=2\pi/\Lambda$ is the supercell momentum, $\Lambda=1950~\mathrm{nm}$ is the metasurface period, and $\theta_i=10^\circ$ is the incidence angle. This relation corresponds directly to the dispersion diagram in Fig.~\hyperref[fig:Fig4]{\ref*{fig:Fig4}(d)}, where each point represents a space--time diffraction channel. In that diagram, $\theta_{mn}$ can be obtained from the ratio of the transverse-momentum coordinate to the frequency coordinate, equivalently from the slope of the line connecting the origin to the point associated with the corresponding diffraction order. The propagating channels lie above the light lines, whereas the gray shaded region below the light lines corresponds to forbidden, evanescent orders. This interpretation follows the generalized space--time grating theory, in which each spatial diffraction order is accompanied by a ladder of temporal harmonics~\cite{taravati2019generalized}.

\begin{figure}
\centering
\includegraphics[scale=1]{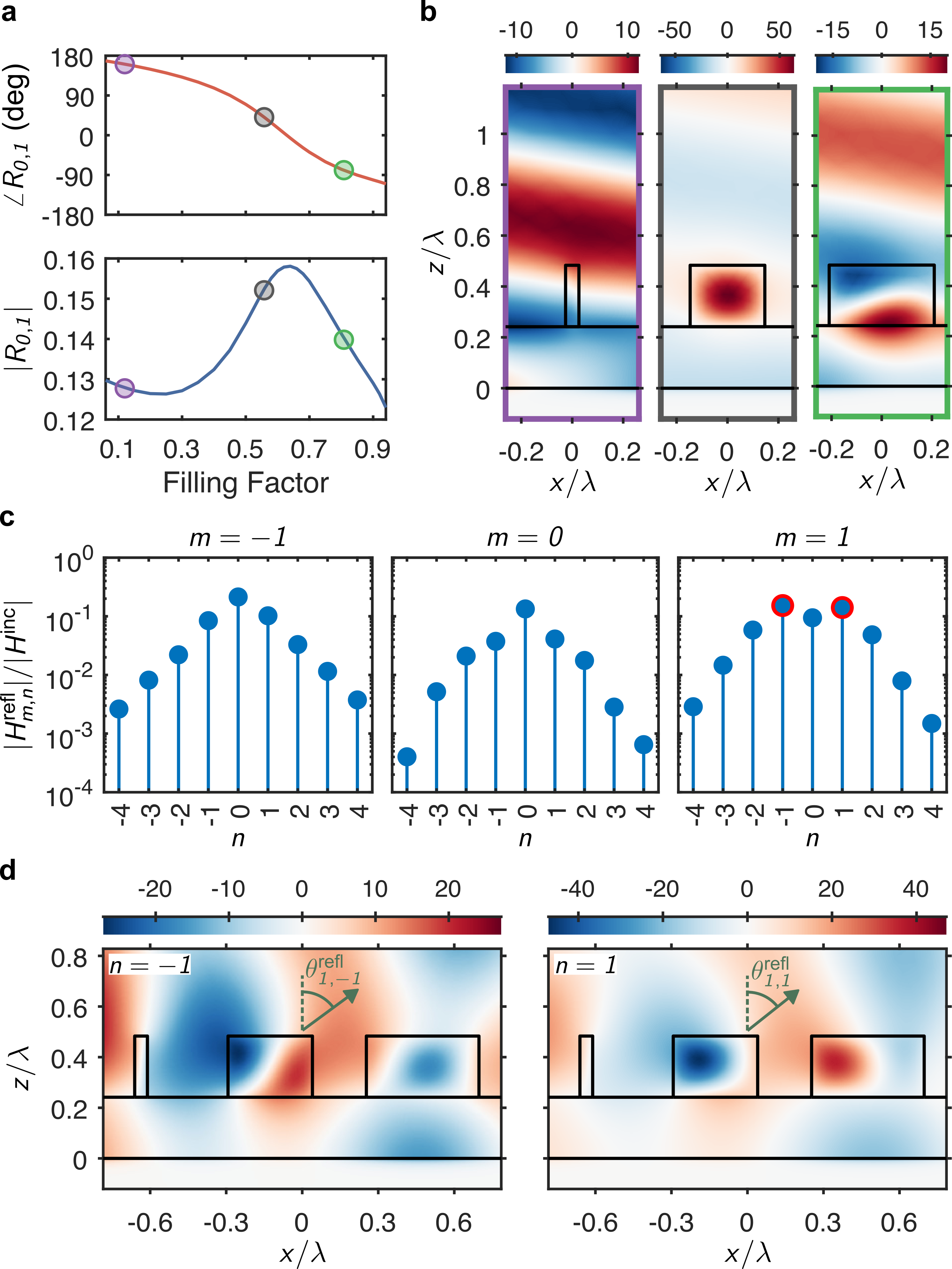}
\caption{
\textbf{Unit cell selection and harmonic-resolved response of the optimized ITO space--time metasurface.}
(a) Phase and amplitude of the unit cell reflection coefficient $R_{0,1}$ for the upper-side harmonic, $n=1$, diffracted into the zeroth spatial order, $m=0$, as a function of dielectric filling factor under TM incidence at $\theta^{\rm inc}=10^\circ$. The colored markers denote the selected unit cells.
(b) Magnetic-field distributions, $H_z$, for the three selected unit cells in panel (a), with frame colors matching the corresponding markers.
(c) Normalized reflected-field amplitudes of the optimized supercell for spatial orders $m=-1,0,1$ and harmonics $n=-4,\ldots,4$, plotted on a common logarithmic scale. The red-circled points indicate the targeted $m=1$ space--time diffraction channels.
(d) Magnetic-field profiles of the optimized supercell at $\omega_{-1}$ and $\omega_{1}$, showing dominant reflection into the $m=1$ order with $\theta^{\rm refl}_{1,-1}=54.9^\circ$ and $\theta^{\rm refl}_{1,1}=53.4^\circ$, respectively. The excitation power is $1~\mathrm{W/m}$ in panels (b,d), and all field maps are in A/m.
}
\label{fig:Fig5}
\end{figure}

The design procedure is summarized in Fig.~\ref{fig:Fig5}. The supercell consists of three unit cells, each of length $650~\mathrm{nm}$, so that the total period is $\Lambda=1950~\mathrm{nm}$. Each individual unit cell is subwavelength and therefore supports only the zeroth propagating spatial order, $m=0$, for the local unit-cell calculation. We sweep the dielectric filling factor of the TiO$_2$ grating and extract the complex reflection coefficient $R_{0,1}$ associated with conversion to the upper sideband, $n=1$, while remaining in the zeroth spatial order. The amplitude and phase of $R_{0,1}$ are shown in Fig.~\hyperref[fig:Fig5]{\ref*{fig:Fig5}(a)}. Because the final device is intended to deflect the frequency-converted field, the phase library is constructed at the sideband frequency rather than at the incident frequency. Three unit-cell geometries are then selected such that their $R_{0,1}$ phases are separated by approximately $120^\circ$, providing a three-level discretization of the desired phase progression across the supercell.

The selected unit cells are shown in Fig.~\hyperref[fig:Fig5]{\ref*{fig:Fig5}(b)} through their magnetic-field distributions. These unit cells provide the initial supercell design, but the final metasurface response is further improved by shape optimization of the TiO$_2$ grating region. In this optimization step, the positions and widths of the dielectric blocks are adjusted while the time-varying ITO layer and gold reflector remain fixed. This refinement is important because the optimized device is not merely a local phase-gradient assembly: near-field coupling between neighboring grating blocks, the dispersive ENZ response of ITO, the gold-backed reflective cavity, and the inter-harmonic coupling all contribute to the final scattering amplitudes. In the Supporting Information, we further show the normalized amplitudes of the deflected $m=1$ sideband beams, $n=-1$ and $n=1$, as functions of the optimization iteration; see Fig.~S3(a). These curves demonstrate how the targeted sideband amplitudes increase through the gradient-based optimization. Because the ITO modulation is harmonic, the two first-order sidebands are generated by the same sinusoidal perturbation of the Drude response. More specifically, for a real single-tone modulation, the coefficients $A_{+1}$ and $A_{-1}$ are complex conjugates. Consequently, the optimization trajectories of the $m=1,n=-1$ and $m=1,n=1$ diffraction amplitudes are strongly correlated, which allows both sideband channels to be enhanced simultaneously without introducing a competing design objective.

Figure~\hyperref[fig:Fig5]{\ref*{fig:Fig5}(c)} shows the harmonic-resolved reflected-field amplitudes of the optimized supercell for spatial orders $m=-1,0,1$ and temporal harmonics $n=-4,\ldots,4$. The amplitudes are plotted on a common logarithmic scale, revealing how the incident power is redistributed among multiple space--time diffraction channels. The red-circled points in Fig.~\hyperref[fig:Fig5]{\ref*{fig:Fig5}(c)} correspond to the red-circled target channels in the dispersion diagram of Fig.~\hyperref[fig:Fig4]{\ref*{fig:Fig4}(d)}. At both sideband frequencies $\omega_{-1}$ and $\omega_1$, the reflected amplitude in the $m=1$ spatial diffraction order is larger than the amplitudes of the other spatial orders. Thus, the sideband fields are effectively redirected into the $m=1$ diffraction channel. This is the central frequency-converting beam-deflection mechanism of the metasurface: the incident field at $\omega_0$ is converted to sideband harmonics and, at those sideband frequencies, preferentially diffracted into the first spatial order.

For the chosen parameters, the generalized grating relation predicts reflection angles of approximately $\theta^{\rm refl}_{1,-1}=54.9^\circ$ and $\theta^{\rm refl}_{1,1}=53.4^\circ$, in agreement with the field profiles shown in Fig.~\hyperref[fig:Fig5]{\ref*{fig:Fig5}(d)}. These two angles are close because the modulation frequency is much smaller than the optical carrier frequency: $\Omega/\omega_0\approx0.012$ for $\lambda_0=1240~\mathrm{nm}$ and $\Omega/2\pi=3~\mathrm{THz}$. The same point is evident in the dispersion diagram of Fig.~\hyperref[fig:Fig4]{\ref*{fig:Fig4}(d)}, where the vertical axis is normalized as $\omega/\omega_0$; adjacent temporal harmonics are separated by only a small vertical distance relative to the optical carrier frequency. As a result, the $m=1$ channels at $\omega_{-1}$ and $\omega_1$ have very similar transverse-momentum-to-frequency ratios and therefore very similar reflection angles.

This example demonstrates the ability of the harmonic-balance framework to analyze and design optical time-varying metasurfaces containing dispersive volumetric media. Around the ENZ frequency, the material response is inherently dispersive because the real part of the permittivity crosses zero with a large spectral slope, and the wave impedance can vary strongly with frequency. Therefore, modulation of the plasma frequency in a Drude-type ENZ medium represents modulation of a dispersive material response, rather than a simple frequency-independent perturbation of the permittivity. By solving all retained Floquet harmonics simultaneously, the proposed framework naturally captures the combined effects of ENZ dispersion, time-dependent Drude coupling, and spatial diffraction from the patterned grating. The resulting design illustrates how dispersive temporal modulation and spatial structuring can be combined to realize frequency-converting beam steering in a realistic optical metasurface platform.

\section{Conclusion}

We have developed a full-wave harmonic-balance framework for dispersive
time-varying photonic structures. The formulation recasts the steady-state
response of a time-periodic electromagnetic system as a coupled set of
frequency-domain full-wave problems in Floquet space. By representing the
time-varying components through induced secondary sources, namely volumetric
polarization densities and surface current densities, the method provides a
unified treatment of modulated bulk media and modulated material sheets. In this
representation, material dispersion is evaluated at the physical frequency of
each generated harmonic, while temporal modulation appears as off-diagonal
coupling between harmonics. The resulting framework therefore captures the
combined effects of frequency dispersion, harmonic conversion, parametric energy
exchange, and full-wave spatial scattering without relying on time marching, quasi-static or
weak dispersion approximations.

The validity and versatility of the framework were demonstrated using two representative dispersive time-varying platforms. In the first example, a parametrically pumped rolled graphene cylinder was analyzed in a regime where the Fermi-level modulation couples the carrier to the down-converted Floquet harmonic, and the harmonic-balance results agreed closely with an analytical Floquet scattering solution, confirming that the method accurately resolves the dispersive graphene surface current, the modulation-induced harmonic coupling, and the resulting enhancement of scattering and absorption. In the second example, the framework was applied to an optical space--time metasurface consisting of a patterned TiO$_2$ grating on a time-modulated ITO ENZ layer, where incorporating the dispersive Drude response of ITO directly in the coupled harmonic system enabled the design and optimization of a reflective frequency-converting metasurface that redirects the generated sidebands into a selected spatial diffraction order. These examples show that the proposed harmonic-balance framework can serve not only as an analysis tool, but also as a design platform for active and time-varying photonic devices with realistic material dispersion and complex geometry, as it directly provides harmonic-resolved fields, powers, absorption, and diffraction amplitudes, offering a transparent route for quantifying how modulation redistributes electromagnetic energy among frequency-shifted channels. This capability is essential for emerging space--time photonic systems based on graphene, transparent conducting oxides, ENZ media, resonant metasurfaces, and other strongly dispersive platforms, and the framework therefore provides a general foundation for the full-wave modeling and inverse design of dispersive time-varying structures for frequency conversion, parametric amplification, harmonic-domain beam steering, and dynamic wavefront control.

\medskip
\noindent\textbf{Supporting Information} \par
Supporting information is available for this paper. Correspondence and requests for materials
should be addressed to W.C.

\medskip
\noindent\textbf{Acknowledgements} \par
This work was supported in part by the National Science Foundation (NSF) under Grant No.
DMR-2323909.

\bibliographystyle{naturemag}   
\bibliography{Refs}

\clearpage

\clearpage

\setcounter{section}{0}
\setcounter{figure}{0}
\setcounter{table}{0}
\setcounter{equation}{0}
\setcounter{page}{1}

\renewcommand{\thesection}{S\arabic{section}}
\renewcommand{\thefigure}{S\arabic{figure}}
\renewcommand{\thetable}{S\arabic{table}}
\renewcommand{\theequation}{S\arabic{equation}}

\renewcommand{\theHsection}{SI.\arabic{section}}
\renewcommand{\theHfigure}{SI.\arabic{figure}}
\renewcommand{\theHtable}{SI.\arabic{table}}
\renewcommand{\theHequation}{SI.\arabic{equation}}

\renewcommand{\bibnumfmt}[1]{[S#1]}
\renewcommand{\citenumfont}[1]{S#1}

\begin{center}
  {\LARGE Supporting Information}\\[8pt]
  {\LARGE\bfseries Full-Wave Harmonic Balance Framework for Dispersive Time-Varying Photonic Structures}\\[10pt]
  Mohammad R. Tavakol$^{1}$ and Wenshan Cai$^{1,2,*}$\\[4pt]
  {\small $^{1}$School of Electrical and Computer Engineering, Georgia Institute of Technology, Atlanta, Georgia 30332, USA}\\
  {\small $^{2}$School of Materials Science and Engineering, Georgia Institute of Technology, Atlanta, Georgia 30332, USA}\\
  {\small $^{*}$Email: wcai@gatech.edu}
\end{center}

\vspace{1em}
\addtocontents{toc}{\protect\setcounter{tocdepth}{2}}
\renewcommand{\contentsname}{Contents}
\tableofcontents
\clearpage

\section{Surface current response of Fermi-level-modulated graphene}

Starting from the time-domain constitutive relation for the rolled graphene sheet, whose dispersive response can be expressed via the Drude equation as \citeS{Galiffi2020WoodAnomalies}
\begin{equation}
\frac{dJ_s(t)}{dt}+\gamma J_s(t)=W(t)E_\phi(t),
\label{eq:td_current_relation_cos_drive}
\end{equation}
we now derive the explicit time-domain surface-current response when the tangential electric field is monochromatic,
\begin{equation}
E_\phi(t)=E_0\cos(\omega_0 t).
\label{eq:ephi_cos_drive}
\end{equation}
Here, \(J_s(t)\) is the azimuthal surface-current density, \(\gamma=1/\tau\) is the carrier relaxation rate, and \(W(t)\) is the time-dependent Drude weight of graphene.

For a single-tone modulation of the graphene Fermi level, the Drude weight is written as
\begin{equation}
W(t)=W_0+\Delta W\cos(\Omega t+\alpha),
\label{eq:W_single_tone_time}
\end{equation}
where
\begin{equation}
W_0=\frac{e^2}{\pi\hbar^2}E_{F0},
\qquad
\Delta W=\frac{e^2}{\pi\hbar^2}\Delta E_F .
\label{eq:W0_deltaW_def}
\end{equation}

Substituting \eqref{eq:ephi_cos_drive} and \eqref{eq:W_single_tone_time} into the right-hand side of \eqref{eq:td_current_relation_cos_drive} gives
\begin{equation}
W(t)E_\phi(t)
=
\left[W_0+\Delta W\cos(\Omega t+\alpha)\right]E_0\cos(\omega_0 t).
\label{eq:drive_product_initial}
\end{equation}
Using the trigonometric identity
\begin{equation}
\cos A \cos B
=
\frac{1}{2}\left[\cos(A+B)+\cos(A-B)\right],
\label{eq:cos_product_identity}
\end{equation}
we obtain
\begin{equation}
W(t)E_\phi(t)
=
W_0E_0\cos(\omega_0 t)
+
\frac{\Delta W E_0}{2}
\cos\!\left[(\omega_0+\Omega)t+\alpha\right]
+
\frac{\Delta W E_0}{2}
\cos\!\left[(\omega_0-\Omega)t-\alpha\right].
\label{eq:drive_product_expanded}
\end{equation}
Therefore, the modulation of the Drude weight generates two additional driving frequencies at
\(\omega_0+\Omega\) and \(\omega_0-\Omega\), in addition to the original frequency \(\omega_0\).

To solve \eqref{eq:td_current_relation_cos_drive}, we use the Fourier transform convention
\begin{equation}
\mathcal{F}\{f(t)\}=f(\omega)=\int_{-\infty}^{\infty}f(t)e^{-j\omega t}\,dt,
\qquad
\mathcal{F}^{-1}\{f(\omega)\}
=
\frac{1}{2\pi}
\int_{-\infty}^{\infty}f(\omega)e^{j\omega t}\,d\omega .
\label{eq:FT_convention}
\end{equation}
Using the derivative property
\begin{equation}
\mathcal{F}\left\{\frac{dJ_s(t)}{dt}\right\}
=
j\omega J_s(\omega),
\label{eq:FT_derivative_property}
\end{equation}
the Fourier transform of \eqref{eq:td_current_relation_cos_drive} becomes
\begin{equation}
\left(\gamma+j\omega\right)J_s(\omega)
=
\mathcal{F}\{W(t)E_\phi(t)\}.
\label{eq:FT_current_relation}
\end{equation}
Hence,
\begin{equation}
J_s(\omega)
=
\frac{\mathcal{F}\{W(t)E_\phi(t)\}}{\gamma+j\omega}.
\label{eq:Js_freq_response_general}
\end{equation}

The Fourier transform of a shifted cosine is
\begin{equation}
\mathcal{F}\left\{A\cos(\omega_c t+\theta)\right\}
=
\pi A
\left[
e^{j\theta}\delta(\omega-\omega_c)
+
e^{-j\theta}\delta(\omega+\omega_c)
\right].
\label{eq:FT_cos_shifted}
\end{equation}
Applying \eqref{eq:FT_cos_shifted} to \eqref{eq:drive_product_expanded}, we obtain
\begin{align}
J_s(\omega)
&=
\frac{\pi W_0E_0}
{\gamma+j\omega}
\left[
\delta(\omega-\omega_0)+\delta(\omega+\omega_0)
\right]
\nonumber\\
&\quad
+
\frac{\pi \Delta W E_0}{2}
\frac{1}{\gamma+j\omega}
\left[
e^{j\alpha}\delta(\omega-\omega_0-\Omega)
+
e^{-j\alpha}\delta(\omega+\omega_0+\Omega)
\right]
\nonumber\\
&\quad
+
\frac{\pi \Delta W E_0}{2}
\frac{1}{\gamma+j\omega}
\left[
e^{-j\alpha}\delta(\omega-\omega_0+\Omega)
+
e^{j\alpha}\delta(\omega+\omega_0-\Omega)
\right].
\label{eq:Js_frequency_delta}
\end{align}

Using the sifting property of the delta function,
\begin{equation}
f(\omega)\delta(\omega-\omega_c)=f(\omega_c)\delta(\omega-\omega_c),
\label{eq:delta_sifting_property}
\end{equation}
the inverse Fourier transform gives the steady-state current response
\begin{align}
J_s^{\rm ss}(t)
&=
\frac{W_0E_0}{2}
\left[
\frac{e^{j\omega_0 t}}{\gamma+j\omega_0}
+
\frac{e^{-j\omega_0 t}}{\gamma-j\omega_0}
\right]
\nonumber\\
&\quad
+
\frac{\Delta W E_0}{4}
\left[
\frac{e^{j[(\omega_0+\Omega)t+\alpha]}}{\gamma+j(\omega_0+\Omega)}
+
\frac{e^{-j[(\omega_0+\Omega)t+\alpha]}}{\gamma-j(\omega_0+\Omega)}
\right]
\nonumber\\
&\quad
+
\frac{\Delta W E_0}{4}
\left[
\frac{e^{j[(\omega_0-\Omega)t-\alpha]}}{\gamma+j(\omega_0-\Omega)}
+
\frac{e^{-j[(\omega_0-\Omega)t-\alpha]}}{\gamma-j(\omega_0-\Omega)}
\right].
\label{eq:Js_time_complex_pair}
\end{align}

Equation \eqref{eq:Js_time_complex_pair} is real-valued because each positive-frequency term is accompanied by its complex-conjugate negative-frequency term. It can therefore be written in the compact real form
\begin{align}
J_s^{\rm ss}(t)
&=
\frac{W_0E_0}{\sqrt{\gamma^2+\omega_0^2}}
\cos\!\left[
\omega_0 t-\tan^{-1}\!\left(\frac{\omega_0}{\gamma}\right)
\right]
\nonumber\\
&\quad
+
\frac{\Delta W E_0}
{2\sqrt{\gamma^2+(\omega_0+\Omega)^2}}
\cos\!\left[
(\omega_0+\Omega)t+\alpha
-\tan^{-1}\!\left(\frac{\omega_0+\Omega}{\gamma}\right)
\right]
\nonumber\\
&\quad
+
\frac{\Delta W E_0}
{2\sqrt{\gamma^2+(\omega_0-\Omega)^2}}
\cos\!\left[
(\omega_0-\Omega)t-\alpha
-\tan^{-1}\!\left(\frac{\omega_0-\Omega}{\gamma}\right)
\right].
\label{eq:Js_time_real_final}
\end{align}

The complete time-domain solution also includes the homogeneous response of
\eqref{eq:td_current_relation_cos_drive},
\begin{equation}
J_s^{\rm hom}(t)=C e^{-\gamma t},
\label{eq:Js_homogeneous}
\end{equation}
where \(C\) is determined by the initial condition. Thus,
\begin{equation}
J_s(t)=J_s^{\rm ss}(t)+C e^{-\gamma t}.
\label{eq:Js_total_time_response}
\end{equation}
For the steady-state response, or after a time much longer than the relaxation time
\(\tau=1/\gamma\), the transient term vanishes and the current is given by
\eqref{eq:Js_time_real_final}.

This result shows explicitly that a monochromatic field at \(\omega_0\) produces surface-current harmonics at
\begin{equation}
\omega_0,\qquad \omega_0+\Omega,\qquad \omega_0-\Omega.
\label{eq:generated_current_frequencies}
\end{equation}
The sideband currents originate from the temporal modulation of the graphene Drude weight, which mixes the incident electric-field frequency with the modulation frequency.

\section{Floquet scattering from a graphene cylinder with modulated Fermi level}
\label{sec:floquet_cylinder_graphene}

We consider two-dimensional TM$_y$ scattering from an infinite dielectric cylinder of radius $a$, invariant along the $y$-direction ($\partial/\partial y = 0$). Region~I ($r>a$) is free space with permittivity $\varepsilon_{\mathrm I}=\varepsilon_0$, and Region~II ($r<a$) is a homogeneous dielectric with permittivity $\varepsilon_{\mathrm{II}}=\varepsilon_1$. All media are non-magnetic, with permeability $\mu_0$. The cylindrical coordinates $(r,\phi)$ are defined in the $x$--$z$ plane as $x=r\cos\phi$, $z=r\sin\phi$, where $\phi=0$ corresponds to the $+x$ direction and $\phi=\pi/2$ to the $+z$ direction. Throughout, we adopt the electrical-engineering time convention $e^{j\omega t}$. A conductive coating is modeled as a zero-thickness sheet at $r=a$ with a time-varying surface response that is uniform in $\phi$.

\paragraph{Time-domain constitutive relation for graphene.}
Rather than introducing a conductivity as a function of frequency and time simultaneously, we describe the graphene sheet via a time-domain constitutive relation that links the surface current density to the tangential electric field. In the intraband (Drude-like) regime \citeS{Hanson2008Dyadic,Falkovsky2007Space}, the graphene surface current density $J_s(t)$ (tangential to the sheet) satisfies the first-order differential equation
\begin{equation}
\frac{d J_s(t)}{dt}
+
\frac{1}{\tau}\,J_s(t)
=
W(t)\,E_t(t),
\label{eq:time_domain_drude_sheet}
\end{equation}
where $\tau$ is the carrier relaxation time, $E_t(t)$ is the tangential electric field at the sheet (here $E_t\equiv E_\phi(r=a,\phi,t)$ for the present cylindrical geometry), and $W(t)$ is the (possibly time-dependent) Drude weight.

The Drude weight is related to the graphene Fermi level $E_F(t)$ through
\begin{equation}
W(t)\triangleq\frac{e^2}{\pi\hbar^2}\,E_F(t).
\label{eq:drude_weight_def}
\end{equation}

\paragraph{Temporal modulation of the Fermi level.}
We assume that the Fermi level is harmonically modulated in time as
\begin{equation}
E_F(t)=E_{F0}+\Delta E_F\cos(\Omega t+\alpha),
\label{eq:EF_modulation_main}
\end{equation}
where $\Omega$ is the modulation angular frequency and $\alpha$ is the modulation phase. Consequently, the Drude weight becomes periodic and admits a Fourier-series representation
\begin{equation}
W(t)=\sum_{m\in\mathbb{Z}}W_m e^{jm\Omega t}.
\label{eq:W_series_general}
\end{equation}
For the single-tone modulation \eqref{eq:EF_modulation_main}, only $m=0,\pm1$ are nonzero and one can write
\begin{equation}
W(t)=W_0+W_{+1}e^{j\Omega t}+W_{-1}e^{-j\Omega t},
\label{eq:W_fourier_main}
\end{equation}
with Fourier coefficients
\begin{equation}
W_0=\frac{e^2}{\pi\hbar^2}E_{F0},\qquad
W_{\pm1}=\frac{e^2}{\pi\hbar^2}\frac{\Delta E_F}{2}e^{\pm j\alpha}.
\label{eq:W_coeff_main}
\end{equation}

\paragraph{Floquet harmonics and wavenumbers.}
Under time-periodic excitation at modulation frequency $\Omega$, the fields develop Floquet harmonics at
\begin{equation}
\omega_n=\omega+n\Omega,\qquad n\in\mathbb{Z}.
\end{equation}
We truncate the Floquet index to $n=-N,\dots,N$, and the azimuthal order to $m=-M,\dots,M$. The wavenumbers in the two regions for each Floquet harmonic are
\begin{equation}
k_{\mathrm{I},n}=\omega_n\sqrt{\mu_0\varepsilon_0},\qquad
k_{\mathrm{II},n}=\omega_n\sqrt{\mu_0\varepsilon_1}.
\end{equation}

For TM$_y$ polarization, $\mathbf H=\hat{\mathbf y}H_y$, and at each Floquet frequency $\omega_n$ the tangential electric field satisfies
\begin{equation}
E_{\phi,n}^{\mathrm I,\mathrm{II}}(r,\phi)
=
\frac{1}{j\omega_n\varepsilon_{\mathrm I,\mathrm{II}}}\,
\frac{\partial H_{y,n}^{\mathrm I,\mathrm{II}}(r,\phi)}{\partial r}.
\end{equation}
The incident field is chosen as a plane wave propagating toward $-z$,
\begin{equation}
H_y^{\mathrm{inc}}(r,\phi,t)=H_0\,e^{+jk_{\mathrm{I},0}z}\,e^{j\omega t}
=H_0\,e^{+jk_{\mathrm{I},0}r\sin\phi}\,e^{j\omega t},
\end{equation}
which admits the angular expansion \citeS{balanis2012advanced}
\begin{equation}
e^{jk r\sin\phi}=\sum_{m=-\infty}^{\infty}J_m(kr)e^{jm\phi}.
\end{equation}
Hence, the incident field contains all azimuthal orders $m$ but only the carrier harmonic $n=0$. Because the geometry is circular and the sheet is uniform in $\phi$, distinct azimuthal orders remain decoupled; the plane wave simply provides a known forcing in each $m$-channel.

We expand the fields as sums of azimuthal harmonics. In Region~I ($r>a$), we represent outgoing radiation with Hankel functions of the second kind $H_m^{(2)}$, consistent with the $e^{j\omega t}$ convention. The Floquet-domain magnetic field in Region~I is written as
\begin{equation}
H_{y,n}^{\mathrm I}(r,\phi)=
\sum_{m=-M}^{M}
\left[
\delta_{n0}H_0\,J_m(k_{\mathrm{I},n}r)
+
H_0\,c_{m,n}\,H_m^{(2)}(k_{\mathrm{I},n}r)
\right]e^{jm\phi}.
\end{equation}
The coefficients $c_{m,n}$ are unknown and represent the amplitudes of the scattered Floquet harmonics. In Region~II ($r<a$), regularity at the origin gives
\begin{equation}
H_{y,n}^{\mathrm{II}}(r,\phi)=
\sum_{m=-M}^{M}
H_0\,d_{m,n}\,J_m(k_{\mathrm{II},n}r)\,e^{jm\phi},
\end{equation}
where $d_{m,n}$ are internal field coefficients.

\paragraph{Surface current harmonics from the graphene constitutive law.}
At the sheet $r=a$, the tangential electric field is $E_t(t)\equiv E_\phi(a,\phi,t)$ and the sheet current density is $J_s(t)\equiv J_s(\phi,t)$. Because the Drude weight is periodic, we expand the tangential electric field and sheet current into Floquet harmonics,
\begin{align}
E_t(t)
&=
\sum_{n\in\mathbb{Z}}
E_n(\phi)\,e^{j(\omega+n\Omega)t},
\\
J_s(t)
&=
\sum_{n\in\mathbb{Z}}
J_{s,n}(\phi)\,e^{j(\omega+n\Omega)t},
\end{align}
and likewise expand $W(t)$ as in \eqref{eq:W_series_general}. Substituting these series into the time-domain constitutive relation \eqref{eq:time_domain_drude_sheet} and equating equal time-harmonic terms yields the discrete Floquet relation
\begin{equation}
\left[j(\omega+n\Omega)+\frac{1}{\tau}\right]J_{s,n}(\phi)
=
\sum_{m\in\mathbb{Z}} W_m\,E_{n-m}(\phi).
\label{eq:floquet_current_general}
\end{equation}
For single-tone modulation \eqref{eq:W_fourier_main}, only $m=0,\pm1$ contribute, and \eqref{eq:floquet_current_general} reduces to the nearest-neighbor coupling law
\begin{equation}
\left[j(\omega+n\Omega)+\frac{1}{\tau}\right]J_{s,n}(\phi)
=
W_0\,E_n(\phi)
+
W_{+1}\,E_{n-1}(\phi)
+
W_{-1}\,E_{n+1}(\phi).
\label{eq:floquet_current_nn}
\end{equation}
This expression makes explicit that the temporally modulated graphene couples adjacent Floquet harmonics through the coefficients $W_{\pm1}$, while $W_0$ produces the usual intraband response in the static limit.

At the sheet $r=a$, the electromagnetic boundary conditions at each Floquet index $n$ are (i) continuity of tangential electric field,
\begin{equation}
E_{\phi,n}^{\mathrm I}(a,\phi) = E_{\phi,n}^{\mathrm{II}}(a,\phi),
\end{equation}
and (ii) jump of the tangential magnetic field equal to the surface current harmonic,
\begin{equation}
H_{y,n}^{\mathrm{II}}(a,\phi) - H_{y,n}^{\mathrm I}(a,\phi) = J_{s,n}(\phi).
\end{equation}
Substituting the azimuthal expansions and exploiting the orthogonality of $e^{j m \phi}$, these conditions are enforced independently for each azimuthal order $m$. The temporal modulation couples different Floquet indices $n$ through \eqref{eq:floquet_current_general}--\eqref{eq:floquet_current_nn}, but does not mix different $m$.

To write the resulting equations compactly, we introduce boundary shorthand (evaluated at $r=a$) for each $m$ and $n$:
\begin{equation}
J_{\mathrm{I},n}=J_m(k_{\mathrm{I},n}a),\quad
H_{\mathrm{I},n}=H_m^{(2)}(k_{\mathrm{I},n}a),\quad
J_{\mathrm{II},n}=J_m(k_{\mathrm{II},n}a),
\end{equation}
and likewise derivatives with respect to the argument,
\begin{equation}
J'_{\mathrm{I},n}=\frac{dJ_m(x)}{dx}\Big|_{x=k_{\mathrm{I},n}a},\quad
H'_{\mathrm{I},n}=\frac{dH_m^{(2)}(x)}{dx}\Big|_{x=k_{\mathrm{I},n}a},\quad
J'_{\mathrm{II},n}=\frac{dJ_m(x)}{dx}\Big|_{x=k_{\mathrm{II},n}a}.
\end{equation}
Using $E_{\phi,n}^{\mathrm I,\mathrm{II}}=-(1/(j\omega_n\varepsilon_{\mathrm I,\mathrm{II}}))\partial_r H_{y,n}^{\mathrm I,\mathrm{II}}$, the electric-field continuity condition yields, for each fixed $m$ and each $n$,
\begin{equation}
\frac{k_{\mathrm{I},n}}{\varepsilon_{\mathrm I}}\Big(\delta_{n0}J'_{\mathrm{I},n}+c_{m,n}H'_{\mathrm{I},n}\Big)
=
\frac{k_{\mathrm{II},n}}{\varepsilon_{\mathrm{II}}}\,d_{m,n}\,J'_{\mathrm{II},n}.
\end{equation}
This relation is diagonal in $n$ and allows the internal coefficient $d_{m,n}$ to be eliminated explicitly:
\begin{equation}
d_{m,n}=
\frac{\varepsilon_{\mathrm{II}}k_{\mathrm{I},n}}{\varepsilon_{\mathrm I}k_{\mathrm{II},n}}
\frac{\delta_{n0}J'_{\mathrm{I},n}+c_{m,n}H'_{\mathrm{I},n}}{J'_{\mathrm{II},n}}.
\end{equation}

\paragraph{Closed system for the scattered coefficients.}
Substituting the eliminated coefficient $d_{m,n}$ into the magnetic jump condition provides a closed set of equations in terms of the unknown scattered coefficients $c_{m,n}$. The coupling among Floquet indices is introduced by the graphene constitutive law \eqref{eq:floquet_current_general} (or \eqref{eq:floquet_current_nn} under single-tone modulation), which expresses the surface current harmonics $J_{s,n}(\phi)$ as linear combinations of tangential electric-field harmonics $E_{\phi,n}(a,\phi)$.

For numerical solution and for transparent physical interpretation, it is convenient to write the per-$m$ system as a finite-dimensional matrix equation over the truncated Floquet indices $n=-N,\dots,N$. We define the unknown coefficient vector
\begin{equation}
\underline{c_m}=
\begin{bmatrix}
c_{m,-N} & \cdots & c_{m,N}
\end{bmatrix}^T,
\end{equation}
and the carrier-selection vector
\begin{equation}
\underline{e}=
\begin{bmatrix}
\delta_{-N,0} & \cdots & \delta_{N,0}
\end{bmatrix}^T,
\end{equation}
which encodes the fact that the incident field exists only at $n=0$. We also define diagonal matrices collecting the boundary function values across Floquet harmonics:
\begin{equation}
\underline{\underline{K_{\mathrm I}}}=\mathrm{diag}(k_{\mathrm{I},-N},\dots,k_{\mathrm{I},N}),\quad
\underline{\underline{K_{\mathrm{II}}}}=\mathrm{diag}(k_{\mathrm{II},-N},\dots,k_{\mathrm{II},N}),
\end{equation}
\begin{equation}
\underline{\underline{J_{\mathrm I}}}=\mathrm{diag}(J_{\mathrm{I},-N},\dots,J_{\mathrm{I},N}),\quad
\underline{\underline{H_{\mathrm I}}}=\mathrm{diag}(H_{\mathrm{I},-N},\dots,H_{\mathrm{I},N}),\quad
\underline{\underline{J_{\mathrm{II}}}}=\mathrm{diag}(J_{\mathrm{II},-N},\dots,J_{\mathrm{II},N}),
\end{equation}
\begin{equation}
\underline{\underline{J'_{\mathrm I}}}=\mathrm{diag}(J'_{\mathrm{I},-N},\dots,J'_{\mathrm{I},N}),\quad
\underline{\underline{H'_{\mathrm I}}}=\mathrm{diag}(H'_{\mathrm{I},-N},\dots,H'_{\mathrm{I},N}),\quad
\underline{\underline{J'_{\mathrm{II}}}}=\mathrm{diag}(J'_{\mathrm{II},-N},\dots,J'_{\mathrm{II},N}),
\end{equation}
and the diagonal Floquet-frequency matrix
\begin{equation}
\underline{\underline{\omega}}=\mathrm{diag}(\omega_{-N},\dots,\omega_N).
\end{equation}

\paragraph{Graphene coupling operator in Floquet space.}
The graphene constitutive relation \eqref{eq:floquet_current_general} can be written in a compact operator form by defining the diagonal matrix
\begin{equation}
\underline{\underline{D}}
=
\mathrm{diag}\!\left(j\omega_{-N}+\frac{1}{\tau},\,\dots,\,j\omega_{N}+\frac{1}{\tau}\right)
=
j\,\underline{\underline{\omega}}+\frac{1}{\tau}\,\underline{\underline{I}},
\label{eq:D_def}
\end{equation}
and the Toeplitz matrix $\underline{\underline{W}}$ with elements
\begin{equation}
(\underline{\underline{W}})_{nq}=W_{n-q},
\qquad n,q\in\{-N,\dots,N\}.
\label{eq:W_toeplitz}
\end{equation}
Then \eqref{eq:floquet_current_general} becomes, at each $\phi$,
\begin{equation}
\underline{\underline{D}}\;\underline{J_s}(\phi)=\underline{\underline{W}}\;\underline{E}(\phi),
\qquad\Rightarrow\qquad
\underline{J_s}(\phi)=\underline{\underline{D}}^{-1}\underline{\underline{W}}\;\underline{E}(\phi),
\label{eq:Js_from_E_operator}
\end{equation}
where $\underline{E}(\phi)=[E_{-N}(\phi),\dots,E_{N}(\phi)]^T$ and
\begin{equation}
\underline{J_s}(\phi)=\begin{bmatrix}J_{s,-N}(\phi) & \cdots & J_{s,N}(\phi)\end{bmatrix}^{T}.
\end{equation}
For single-tone modulation, $\underline{\underline{W}}$ is tridiagonal since only $W_0$ and $W_{\pm1}$ are nonzero.

The time modulation therefore enters through the effective Floquet coupling operator
\begin{equation}
\underline{\underline{\sigma_s}}
\triangleq
\underline{\underline{D}}^{-1}\,\underline{\underline{W}},
\label{eq:sigma_s_def_operator}
\end{equation}
which maps the tangential electric-field harmonic vector into the surface-current harmonic vector, i.e., $\underline{J_s}=\underline{\underline{\sigma_s}}\,\underline{E}$. Elementwise,
\begin{equation}
(\underline{\underline{\sigma_s}})_{nq}
=
\frac{W_{n-q}}{j\omega_n+1/\tau},
\qquad n,q\in\{-N,\dots,N\}.
\label{eq:sigma_s_elements}
\end{equation}
This expression is consistent with the static limit: when $W(t)=W_0$, one obtains $(\underline{\underline{\sigma_s}})_{nq}=\big(W_0/(j\omega_n+1/\tau)\big)\delta_{nq}$, which recovers the familiar intraband graphene conductivity for each harmonic.

\paragraph{Final matrix system for the scattered coefficients.}
With the above definition of $\underline{\underline{\sigma_s}}$, the eliminated system for $\underline{c_m}$ can be written as
\begin{equation}
\Big(\underline{\underline{A_m}}+\underline{\underline{\sigma_s}}\;\underline{\underline{C}}\Big)\underline{c_m}=\underline{f_m},
\end{equation}
where $\underline{\underline{A_m}}$ is the effective ``static'' scattering operator for azimuthal order $m$, $\underline{\underline{C}}$ maps scattered-field coefficients into the tangential electric field contribution appearing in the surface current, and $\underline{f_m}$ is the forcing vector driven by the incident plane wave. These quantities are given explicitly by
\begin{equation}
\underline{\underline{A_m}}=
\frac{\varepsilon_{\mathrm{II}}}{\varepsilon_{\mathrm I}}
\underline{\underline{K_{\mathrm I}}}
\underline{\underline{K_{\mathrm{II}}}}^{-1}
\big(\underline{\underline{J'_{\mathrm{II}}}}\big)^{-1}
\underline{\underline{J_{\mathrm{II}}}}
\underline{\underline{H'_{\mathrm I}}}
-
\underline{\underline{H_{\mathrm I}}},
\end{equation}
\begin{equation}
\underline{\underline{C}}=
\frac{1}{j\varepsilon_{\mathrm I}}\;
\underline{\underline{K_{\mathrm I}}}\;
\underline{\underline{\omega}}^{-1}\;
\underline{\underline{H'_{\mathrm I}}},
\end{equation}
\begin{equation}
\underline{f_m}=
\underline{\underline{J_{\mathrm I}}}\,\underline{e}
-
\frac{\varepsilon_{\mathrm{II}}}{\varepsilon_{\mathrm I}}
\underline{\underline{K_{\mathrm I}}}
\underline{\underline{K_{\mathrm{II}}}}^{-1}
\big(\underline{\underline{J'_{\mathrm{II}}}}\big)^{-1}
\underline{\underline{J_{\mathrm{II}}}}
\underline{\underline{J'_{\mathrm I}}}\,\underline{e}
-
\underline{\underline{\sigma_s}}\;
\frac{1}{j\varepsilon_{\mathrm I}}
\underline{\underline{K_{\mathrm I}}}\;
\underline{\underline{\omega}}^{-1}\;
\underline{\underline{J'_{\mathrm I}}}\,\underline{e}.
\end{equation}

To make the per-element structure fully explicit, let the matrix to be inverted for a given $m$ be denoted as $\underline{\underline{S_m}}=\underline{\underline{A_m}}+\underline{\underline{\sigma_s}}\underline{\underline{C}}$. The linear system reads $\sum_{q=-N}^{N}(S_m)_{nq}\,c_{m,q}=(f_m)_n$, with entries
\begin{equation}
(S_m)_{nq}=(A_m)_{nq}+\sum_{p=-N}^{N}(\sigma_s)_{np}\,(C)_{pq}.
\end{equation}
Since $\underline{\underline{A_m}}$ and $\underline{\underline{C}}$ are diagonal in the Floquet index, their elements are
\begin{equation}
(A_m)_{nq}=
\left[
\frac{\varepsilon_{\mathrm{II}}}{\varepsilon_{\mathrm I}}
\frac{k_{\mathrm{I},n}}{k_{\mathrm{II},n}}
\frac{J_{\mathrm{II},n}}{J'_{\mathrm{II},n}}
H'_{\mathrm{I},n}
-
H_{\mathrm{I},n}
\right]\delta_{nq},
\end{equation}
\begin{equation}
(C)_{nq}=
\left[
\frac{1}{j\varepsilon_{\mathrm I}}
\frac{k_{\mathrm{I},n}}{\omega_n}
H'_{\mathrm{I},n}
\right]\delta_{nq}.
\end{equation}
Combining these facts with \eqref{eq:sigma_s_elements} gives the elementwise form
\begin{equation}
(S_m)_{nq}=
\left[
\frac{\varepsilon_{\mathrm{II}}}{\varepsilon_{\mathrm I}}
\frac{k_{\mathrm{I},n}}{k_{\mathrm{II},n}}
\frac{J_{\mathrm{II},n}}{J'_{\mathrm{II},n}}
H'_{\mathrm{I},n}
-
H_{\mathrm{I},n}
\right]\delta_{nq}
+
\left(\frac{W_{n-q}}{j\omega_n+1/\tau}\right)
\left[
\frac{1}{j\varepsilon_{\mathrm I}}
\frac{k_{\mathrm{I},q}}{\omega_q}
H'_{\mathrm{I},q}
\right].
\end{equation}
Likewise, the forcing vector elements $(f_m)_n$ can be written explicitly as
\begin{equation}
(f_m)_n=
J_{\mathrm{I},n}\,\delta_{n0}
-
\left[
\frac{\varepsilon_{\mathrm{II}}}{\varepsilon_{\mathrm I}}
\frac{k_{\mathrm{I},n}}{k_{\mathrm{II},n}}
\frac{J_{\mathrm{II},n}}{J'_{\mathrm{II},n}}
J'_{\mathrm{I},n}
\right]\delta_{n0}
-
\sum_{q=-N}^{N}
\left(\frac{W_{n-q}}{j\omega_n+1/\tau}\right)
\left[
\frac{1}{j\varepsilon_{\mathrm I}}
\frac{k_{\mathrm{I},q}}{\omega_q}
J'_{\mathrm{I},q}
\right]\delta_{q0}.
\end{equation}
Since $\delta_{q0}$ appears in the last term, the sum collapses to a single contribution, yielding
\begin{equation}
(f_m)_n=
J_{\mathrm{I},n}\,\delta_{n0}
-
\left[
\frac{\varepsilon_{\mathrm{II}}}{\varepsilon_{\mathrm I}}
\frac{k_{\mathrm{I},n}}{k_{\mathrm{II},n}}
\frac{J_{\mathrm{II},n}}{J'_{\mathrm{II},n}}
J'_{\mathrm{I},n}
\right]\delta_{n0}
-
\left(\frac{W_{n}}{j\omega_n+1/\tau}\right)
\left[
\frac{1}{j\varepsilon_{\mathrm I}}
\frac{k_{\mathrm{I},0}}{\omega_0}
J'_{\mathrm{I},0}
\right].
\end{equation}

In practice, solving the problem proceeds as follows. For each azimuthal order $m\in[-M,M]$, one assembles the $(2N+1)\times(2N+1)$ linear system $\underline{\underline{S_m}}\underline{c_m}=\underline{f_m}$ using the above elementwise definitions and solves for $\underline{c_m}$. The collection of coefficients $c_{m,n}$ fully determines the scattered field in Region~I and is sufficient to compute far-field observables, including the scattering width. If internal fields are required, the eliminated coefficients $d_{m,n}$ are recovered directly from the algebraic relation obtained from tangential electric-field continuity, using the already computed $c_{m,n}$.

Finally, the scattering width associated with each propagating Floquet harmonic depends only on the scattered coefficients $c_{m,n}$. Under the present normalization, the harmonic-resolved scattering width is
\begin{equation}
\sigma_{\mathrm{sca}}^{(n)}=\frac{4}{k_{\mathrm{I},n}}\sum_{m=-M}^{M}|c_{m,n}|^2,
\end{equation}
and the total scattering width is obtained by summing over all the harmonics,
\begin{equation}
\sigma_{\mathrm{sca}}^{\mathrm{tot}}=\sum_{n=-N}^{N}\sigma_{\mathrm{sca}}^{(n)}.
\end{equation}
This decomposition emphasizes that the time-modulated sheet redistributes scattering among frequency-converted Floquet channels through the Floquet coupling operator $\underline{\underline{\sigma_s}}$, while each azimuthal order $m$ remains an independent scattering channel driven by the plane-wave angular spectrum.

The absorption cross section can be obtained from the time-averaged Joule
dissipation on the time-varying graphene sheet. For the \(n\)-th Floquet
harmonic, the absorbed power per unit length is written as
\begin{equation}
    P_{\mathrm{abs}}^{(n)}
    =
    \frac{1}{2}
    \int_{0}^{2\pi}
    a\,\Re\!\left\{
    E_{\phi,n}(a,\phi) J_{s,n}^{*}(a,\phi)
    \right\}
    d\phi .
    \label{eq:Pabs_n_def}
\end{equation}
Using the azimuthal expansions of the tangential electric field and the
surface current, together with the orthogonality of \(e^{jm\phi}\), this becomes
\begin{equation}
    P_{\mathrm{abs}}^{(n)}
    =
    \pi a
    \sum_{m=-M}^{M}
    \Re\!\left\{
    E_{\phi,m,n}(a) J_{s,m,n}^{*}(a)
    \right\}.
    \label{eq:Pabs_n_modesum}
\end{equation}
Therefore, the harmonic-resolved absorption cross section is
\begin{equation}
    \sigma_{\mathrm{abs}}^{(n)}
    =
    \frac{P_{\mathrm{abs}}^{(n)}}{I_0}
    =
    \frac{\pi a}{I_0}
    \sum_{m=-M}^{M}
    \Re\!\left\{
    E_{\phi,m,n}(a) J_{s,m,n}^{*}(a)
    \right\},
    \label{eq:sigma_abs_n}
\end{equation}
where \(I_0\) is the incident intensity at the carrier frequency. The total
absorption cross section is obtained by summing over the retained Floquet
harmonics,
\begin{equation}
    \sigma_{\mathrm{abs}}
    =
    \sum_{n=-N}^{N}
    \sigma_{\mathrm{abs}}^{(n)} .
    \label{eq:sigma_abs_total}
\end{equation}
After solving the harmonic-balance system for the scattered coefficients, the
boundary tangential electric-field vector for each azimuthal order is obtained
from the same boundary relation used in the matrix formulation. The
corresponding surface-current vector follows from the Floquet admittance
relation of the graphene sheet. The scalar quantities
\(E_{\phi,m,n}(a)\) and \(J_{s,m,n}(a)\) in
\eqref{eq:sigma_abs_n} are the \(n\)-th entries of these vectors.

For numerical evaluation, the quantities appearing in
\eqref{eq:sigma_abs_n} are obtained directly from the same matrices used in the
harmonic-balance system. For each azimuthal order \(m\), after solving
\begin{equation}
    \underline{\underline{S_m}}\,\underline{c_m}
    =
    \underline{f_m},
\end{equation}
the boundary tangential electric-field vector is
\begin{equation}
    \underline{E_{\phi,m}}(a)
    =
    \underline{E^{\mathrm{inc}}_{\phi,m}}(a)
    +
    \underline{\underline{C}}\,\underline{c_m},
    \label{eq:Ephi_boundary_matrix_abs_corrected}
\end{equation}
where
\begin{equation}
    \underline{E^{\mathrm{inc}}_{\phi,m}}(a)
    =
    \frac{1}{j\varepsilon_{\mathrm I}}\,
    \underline{\underline{K_{\mathrm I}}}\,
    \underline{\underline{\omega}}^{-1}\,
    \underline{\underline{J'_{\mathrm I}}}\,
    \underline{e}.
    \label{eq:Ephi_inc_matrix_abs_corrected}
\end{equation}
Here, \(\underline{\underline{C}}\,\underline{c_m}\) is the scattered-field
contribution to the tangential electric field at \(r=a\), with
\begin{equation}
    \underline{\underline{C}}
    =
    \frac{1}{j\varepsilon_{\mathrm I}}\,
    \underline{\underline{K_{\mathrm I}}}\,
    \underline{\underline{\omega}}^{-1}\,
    \underline{\underline{H'_{\mathrm I}}}.
\end{equation}
The corresponding surface-current vector is obtained from the Floquet
surface-conductivity relation
\begin{equation}
    \underline{J_{s,m}}
    =
    \underline{\underline{\sigma_s}}\,
    \underline{E_{\phi,m}}(a).
    \label{eq:Js_matrix_abs_corrected}
\end{equation}
Therefore,
\begin{equation}
    E_{\phi,m,n}(a)
    =
    \left[
    \underline{E^{\mathrm{inc}}_{\phi,m}}(a)
    +
    \underline{\underline{C}}\,\underline{c_m}
    \right]_n ,
    \label{eq:Ephi_mn_from_matrix_corrected}
\end{equation}
and
\begin{equation}
    J_{s,m,n}(a)
    =
    \left[
    \underline{\underline{\sigma_s}}
    \left(
    \underline{E^{\mathrm{inc}}_{\phi,m}}(a)
    +
    \underline{\underline{C}}\,\underline{c_m}
    \right)
    \right]_n .
    \label{eq:Js_mn_from_matrix_corrected}
\end{equation}
Substituting these entries into the harmonic-resolved absorption expression
gives
\begin{equation}
    \sigma_{\mathrm{abs}}^{(n)}
    =
    \frac{\pi a}{I_0}
    \sum_{m=-M}^{M}
    \Re\!\left\{
    \left[
    \underline{E^{\mathrm{inc}}_{\phi,m}}(a)
    +
    \underline{\underline{C}}\,\underline{c_m}
    \right]_n
    \left[
    \underline{\underline{\sigma_s}}
    \left(
    \underline{E^{\mathrm{inc}}_{\phi,m}}(a)
    +
    \underline{\underline{C}}\,\underline{c_m}
    \right)
    \right]_n^{*}
    \right\}.
    \label{eq:sigma_abs_matrix_corrected}
\end{equation}

\section{Complementary results for the pumped rolled graphene cylinder}
\label{sec:complementary_pumped_graphene_cylinder}

\begin{figure}[!t]
\centering
\includegraphics[scale=0.92]{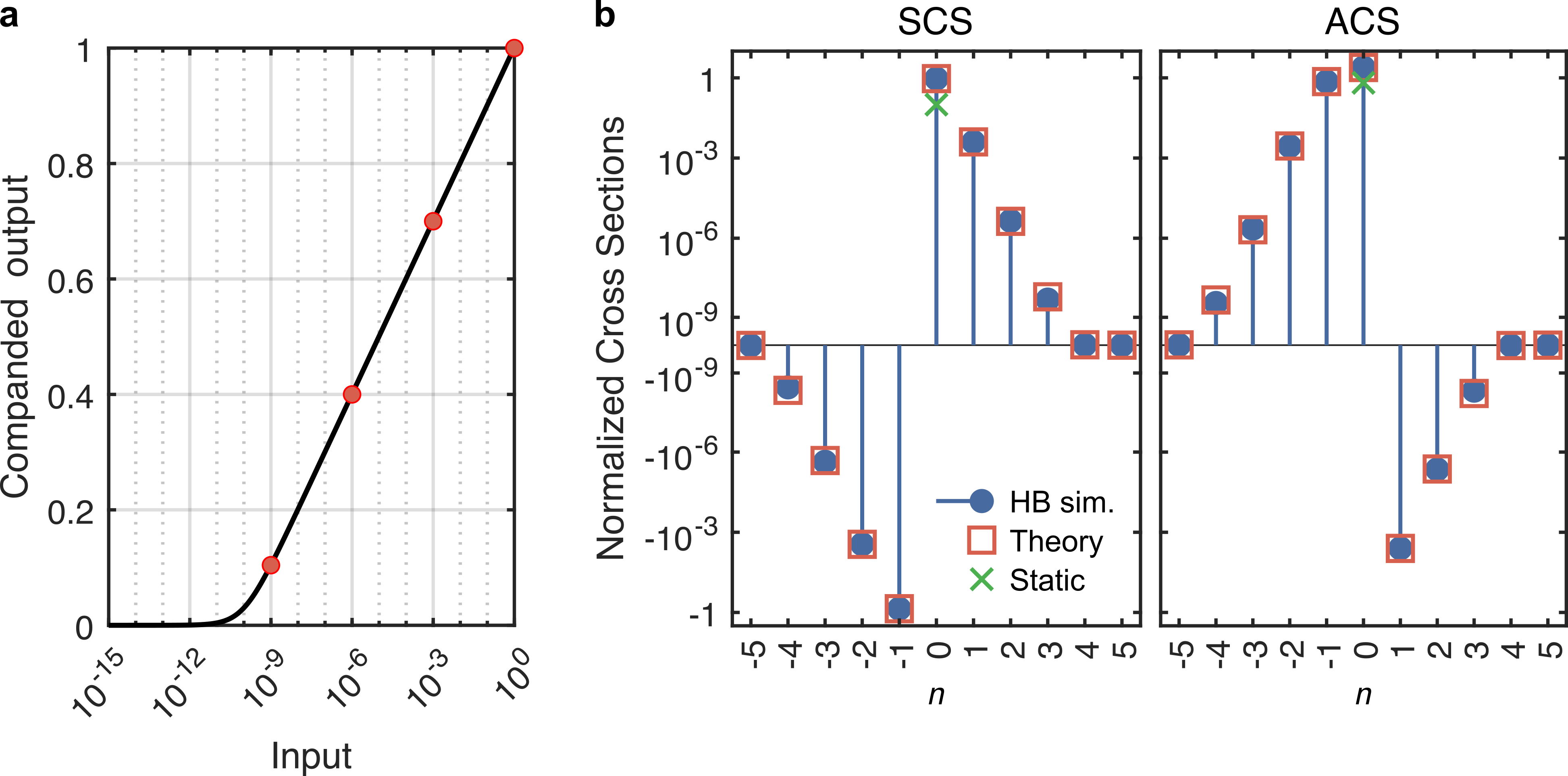}
\caption{
\textbf{Companded harmonic-resolved response of the parametrically pumped rolled graphene cylinder.}
(a) Positive-input $\mu$-law compander characteristic for $\mu=10^{10}$, used to map cross-section values spanning many decades onto a visually resolved nonlinear scale. The red markers correspond to the original values used as vertical-axis ticks in panel (b). 
(b) Normalized scattering and absorption cross sections resolved over Floquet harmonics and plotted on the companded scale. The tick labels indicate the original uncompanded values, making weak harmonic responses down to $\sim 10^{-10}$ distinguishable. The harmonic-balance simulation and analytical theory remain in excellent agreement across all resolved harmonics, with the static response shown for reference.
}
\label{fig:FigS1}
\end{figure}

To further validate the harmonic-balance formulation for the Fermi-level-modulated
rolled graphene cylinder, we provide additional harmonic-resolved scattering and
absorption results in Fig.~\ref{fig:FigS1}. In the main text, the dominant
harmonics are sufficient to describe the principal frequency-conversion response.
However, many higher-order Floquet harmonics have cross sections that are several
orders of magnitude smaller than the main response and therefore cannot be clearly
distinguished on a conventional linear scale. To make these weak harmonic
channels visible, we plot the normalized scattering and absorption cross sections
using a $\mu$-law companded scale.

The $\mu$-law compression function is defined as
\begin{equation*}
    F(x)
    =
    \operatorname{sgn}(x)
    \frac{\log\left(1+\mu |x|\right)}
    {\log\left(1+\mu\right)},
\end{equation*}
with the corresponding inverse mapping
\begin{equation*}
    F^{-1}(y)
    =
    \operatorname{sgn}(y)
    \frac{1}{\mu}
    \left[
    \left(1+\mu\right)^{|y|}-1
    \right].
\end{equation*}
This nonlinear mapping preserves the sign of the input while compressing a large
dynamic range into a finite visual interval. Therefore, very small cross-section
values, including those associated with harmonics far from the dominant harmonic
channels, become distinguishable without obscuring the stronger response. As
shown in Fig.~\ref{fig:FigS1}, the harmonic-balance results remain in excellent
agreement with the analytical Floquet scattering formulation developed in
Sec.~\ref{sec:floquet_cylinder_graphene} across the full set of resolved harmonics. In
particular, the agreement persists even for weak harmonic responses with
normalized cross sections approaching $10^{-10}$, demonstrating that the proposed
harmonic-balance framework accurately captures both the dominant and the highly
suppressed frequency-converted channels.

\begin{figure}[!t]
\centering
\includegraphics[scale=0.92]{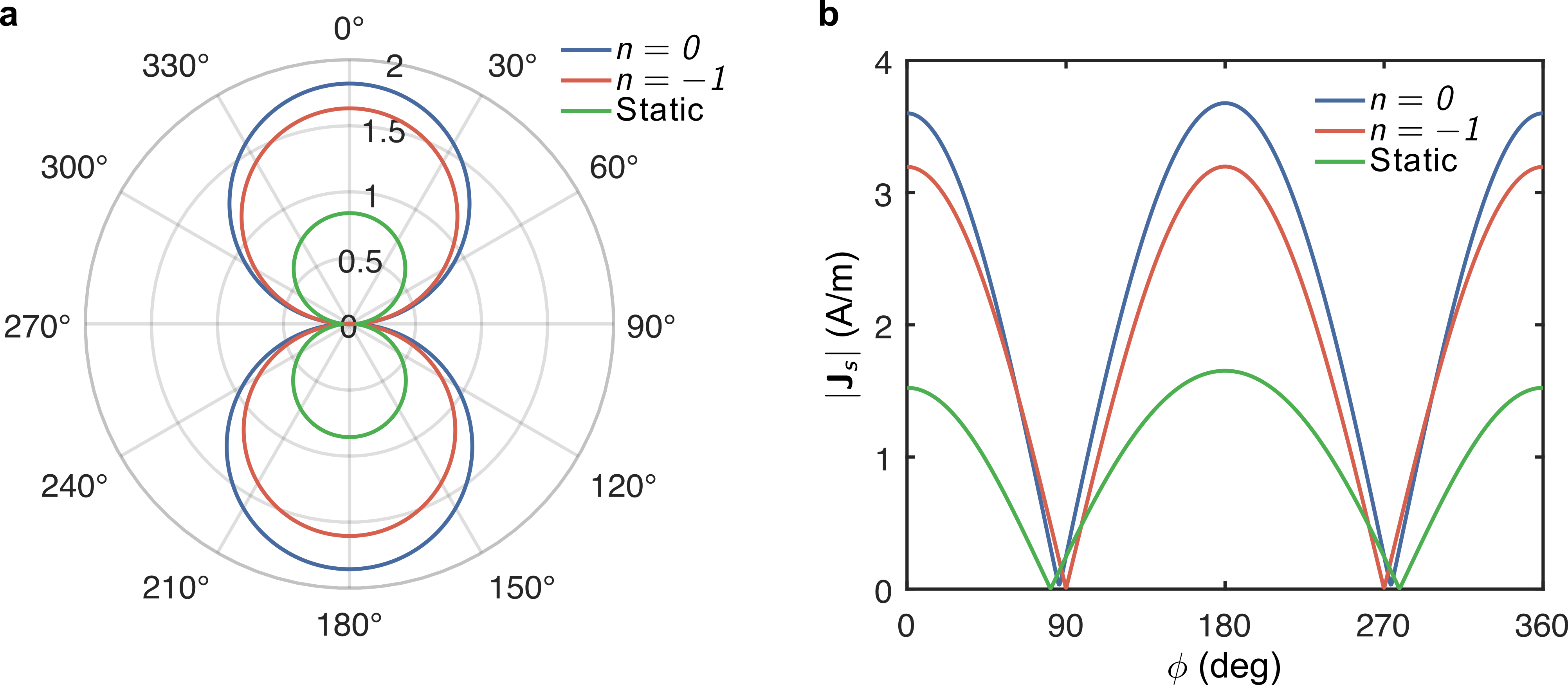}
\caption{
\textbf{Far-field radiation and surface-current profiles of the parametrically pumped rolled graphene cylinder.}
(a) Far-field electric-field pattern, $\mathbf{E}^{\rm far}(\phi)$, for the static cylinder and for the $n=0$ and $n=-1$ Floquet harmonics of the parametrically pumped graphene cylinder. 
(b) Graphene surface-current density versus polar angle $\phi$ for the same cases, comparing the static response with the carrier and down-converted harmonic currents generated by the temporal modulation.
}
\label{fig:FigS2}
\end{figure}

Figure~\ref{fig:FigS2} provides additional field-level insight into the same
pumped rolled graphene cylinder configuration. Specifically,
Fig.~\hyperref[fig:FigS2]{\ref*{fig:FigS2}(a)} compares the far-field electric-field
pattern of the static cylinder with the dominant Floquet harmonics of the
modulated structure. For the modulated case, only the $n=0$ and $n=-1$ harmonics
are shown because these channels dominate the scattered response. The comparison
illustrates how temporal modulation redistributes the scattered field among
Floquet harmonics and can enhance the far-field amplitude relative to the
corresponding static configuration.

The same modulation-induced enhancement is also evident in the graphene surface
current distribution shown in Fig.~\hyperref[fig:FigS2]{\ref*{fig:FigS2}(b)}.
The surface current density is plotted as a function of the polar angle $\phi$,
again comparing the static case with the $n=0$ and $n=-1$ harmonics of the pumped
structure. These current profiles clarify the physical origin of the far-field
response: modulation of the graphene Fermi level modifies the induced surface
current on the rolled sheet, and the resulting harmonic surface currents act as
sources for the frequency-converted scattered fields. Overall, Fig.~\ref{fig:FigS1} serves as a stringent accuracy benchmark for the
proposed harmonic-balance framework by comparing it with the analytical Floquet
scattering formulation over a large dynamic range. In contrast,
Fig.~\ref{fig:FigS2} provides complementary field- and current-level results,
showing that the far-field and induced surface current responses are consistent
with the scattering and absorption cross section trends reported for the pumped
rolled graphene cylinder.

\section{Shape Optimization of the ITO-Based Space--Time Metasurface}
\label{sec:SM_optimization}

The supercell obtained from the unit-cell phase library provides only an
approximate realization of the desired space--time response, because it neglects
near-field coupling between adjacent grating blocks, the dispersive
epsilon-near-zero response of the modulated ITO layer, and the inter-harmonic
coupling that redistributes power among the generated sidebands. To refine the
initial design, we apply a gradient-based shape optimization directly to the
full harmonic-balance model, so that all retained Floquet harmonics are solved
simultaneously at every design iteration.

The design freedom is restricted to the TiO$_2$ grating layer. Each of the three
dielectric blocks in the supercell is allowed to translate along $x$ and to be
scaled in width, while its thickness, the ITO layer, the gold reflector, and the
supercell period $\Lambda$ remain fixed. This yields a compact design vector
$\boldsymbol{\xi}\in\mathbb{R}^{6}$ containing the lateral position and width of
each block. Bounds are imposed on the translations and widths so that
neighboring blocks cannot overlap or cross the periodic boundary. Rather than
regenerating the geometry at each iteration, the design update is applied as a
smooth deformation of the computational mesh, which keeps the mesh connectivity
fixed and allows the design sensitivities to be evaluated analytically.

The device targets the frequency-converted first spatial diffraction order in
reflection. Let $H^{\rm refl}_{m,n}$ denote the complex amplitude of the
reflected magnetic field projected onto the Floquet--Bloch channel with spatial
order $m$ and temporal harmonic $n$, obtained by overlap integration of the
harmonic-balance solution on the port plane. The optimization maximizes the
power carried by the two targeted sideband channels,
\begin{equation}
    Q(\boldsymbol{\xi})
    =
    \sum_{n=\pm 1}
    \frac{\left| H^{\rm refl}_{1,n} \right|^{2}}
         {\left| H^{\rm inc} \right|^{2}} ,
    \label{eq:SM_objective}
\end{equation}
where $H^{\rm inc}$ is the incident magnetic-field amplitude. The objective is
formulated in terms of the squared modulus rather than the modulus itself, since
the former is a smooth function of the complex field unknowns and therefore
admits well-defined partial derivatives; maximizing either quantity yields the
same optimal design. Note that $Q$ deliberately rewards both the upper and lower
sidebands, $n=+1$ and $n=-1$, at the same spatial order $m=1$, consistent with
the frequency-converting beam-deflection functionality described in the main
text.

Writing the discretized harmonic-balance system as
$\mathcal{L}(\underline{\mathbf{X}},\boldsymbol{\xi})=\mathbf{0}$, with
$\underline{\mathbf{X}}$ the multi-harmonic field vector and
$\mathbf{J}=\partial\mathcal{L}/\partial\underline{\mathbf{X}}$ the system
Jacobian, the gradient of the objective is obtained by the adjoint method,
\begin{equation}
    \mathbf{J}^{\mathsf{T}}\,\underline{\mathbf{X}}^{*}
    =
    \left(
        \frac{\partial Q}{\partial \underline{\mathbf{X}}}
    \right)^{\!\mathsf{T}} ,
    \qquad
    \frac{\mathrm{d} Q}{\mathrm{d} \boldsymbol{\xi}}
    =
    \frac{\partial Q}{\partial \boldsymbol{\xi}}
    -
    \left( \underline{\mathbf{X}}^{*} \right)^{\!\mathsf{T}}
    \frac{\partial \mathcal{L}}{\partial \boldsymbol{\xi}} .
    \label{eq:SM_adjoint}
\end{equation}
The key practical point is that the finite-element framework assembles and
factorizes $\mathbf{J}$ during the forward harmonic-balance solve. The adjoint
field $\underline{\mathbf{X}}^{*}$ is therefore recovered by a single
back-substitution with the transposed factor, and the full gradient with respect
to all design variables follows at negligible additional cost. Unlike explicit
time-domain solvers, which require a separate reverse-time simulation to
construct the adjoint field, no second electromagnetic simulation is needed
here. Because the modulated ITO layer couples the harmonics through the
time-dependent Drude response, the adjoint solve automatically accounts for the
sensitivity of every retained harmonic to a perturbation of the grating
geometry. The design is updated using a globally convergent variant of the
method of moving asymptotes, which is well suited to smooth, bound-constrained
problems with a small number of design variables, and a per-iteration move limit
is enforced to prevent excessive distortion of the deforming mesh. Starting from
the phase-library supercell, the optimization is run for $50$ iterations.

\begin{figure}[!t]
\centering
\includegraphics[scale=0.92]{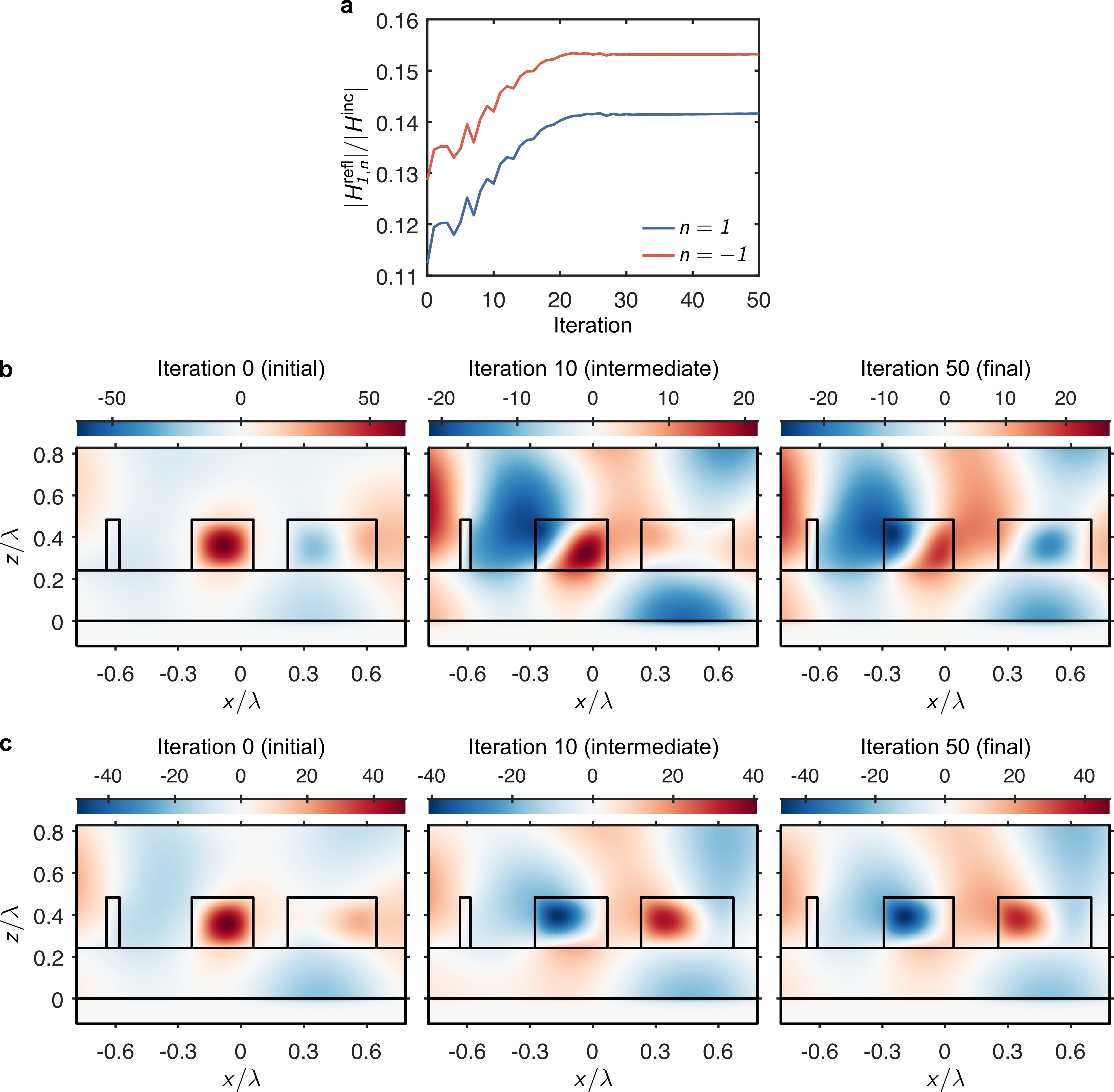}
\caption{
\textbf{Evolution of the targeted space--time diffraction channels during shape optimization.}
(a) Normalized reflected amplitudes of the $m=1$ spatial diffraction channel at the lower and upper sideband harmonics, $n=-1$ and $n=1$, as a function of shape-optimization iteration. Both targeted channels increase during the optimization process.
(b,c) Magnetic-field distributions, $H_z$, at representative optimization iterations for the lower sideband, $\omega_{-1}$, and upper sideband, $\omega_{1}$, respectively, under an incident power of $1~\mathrm{W/m}$ over one supercell period. The field profiles show the gradual formation of the desired reflected diffraction response from the initial geometry to the final optimized metasurface. Color bars in panels (b,c) are in units of A/m.
}
\label{fig:FigS3}
\end{figure}

In Fig.~\hyperref[fig:FigS3]{\ref*{fig:FigS3}(a)}, the normalized reflected
amplitudes $|H^{\rm refl}_{1,n}|/|H^{\rm inc}|$ of both sidebands increase
monotonically, apart from small oscillations in the early iterations. The upper
sideband ($n=1$) rises from approximately $0.112$ to $0.142$ and the lower
sideband ($n=-1$) from approximately $0.130$ to $0.153$, corresponding to
relative improvements of roughly $27\%$ and $18\%$, respectively. Both curves
plateau after about $25$--$30$ iterations, indicating that the design has
converged to a local optimum within the imposed bounds. That both channels
improve simultaneously confirms that the two sidebands are not competing for the
same power budget; instead, the optimizer redistributes power away from the
untargeted $m=0$ and $m=-1$ channels.

The corresponding geometric changes are modest. Measured from the center of the
supercell, the initial phase-library design places the three blocks at
$x$-centers of $-760.0$, $-110.0$, and $540.0~\mathrm{nm}$, with widths of
$78.0$, $364.0$, and $526.5~\mathrm{nm}$, equivalent to filling factors of
$0.12$, $0.56$, and $0.81$ of the $650~\mathrm{nm}$ unit cell. After $50$
iterations the blocks are centered at $-789.6$, $-156.8$, and
$590.0~\mathrm{nm}$, with widths of $62.4$, $413.6$, and $551.7~\mathrm{nm}$,
corresponding to filling factors of $0.096$, $0.636$, and $0.849$. The narrow
first block is thus made narrower still and displaced outward, whereas the two
wider blocks are broadened, and no block is displaced by more than
$50~\mathrm{nm}$, i.e.\ less than $3\%$ of the supercell period. Most of this
rearrangement occurs early: by iteration $10$ the blocks have already reached
widths of $62.4$, $428.7$, and $547.6~\mathrm{nm}$ and centers of $-757.2$,
$-128.9$, and $559.2~\mathrm{nm}$, so that the remaining forty iterations refine
the design only slightly while the targeted amplitudes in
Fig.~\hyperref[fig:FigS3]{\ref*{fig:FigS3}(a)} continue to grow. This indicates
that the performance gain originates from fine adjustment of the inter-block
near-field coupling rather than from a gross redistribution of the dielectric
material, and it justifies treating the phase-library supercell as a good
initial guess.

Figures~\hyperref[fig:FigS3]{\ref*{fig:FigS3}(b)} and
\hyperref[fig:FigS3]{\ref*{fig:FigS3}(c)} show the magnetic-field distributions
$H_z$ at the lower and upper sideband frequencies, $\omega_{-1}$ and
$\omega_{1}$, for these three representative iterations. At iteration $0$ the
sideband fields are weak and largely localized within the dielectric blocks,
reflecting the fact that the phase-library supercell was assembled from isolated
unit-cell responses. As the optimization proceeds, the block positions and
widths adjust so that the sideband fields extend into the ENZ region and acquire
the transverse phase progression required for deflection into the $m=1$ order.
By iteration $50$, both sidebands display the tilted wavefronts characteristic
of the intended frequency-converting deflection, in agreement with the far-field
diffraction angles reported in the main text.


\bibliographystyleS{naturemag}
\bibliographyS{RefsS}

\clearpage

\end{document}